\documentclass[12pt]{article}
\RequirePackage[ colorlinks=true,urlcolor=blue,anchorcolor=blue,citecolor=blue,filecolor=blue,               linkcolor=blue,menucolor=blue,linktocpage=true,pdfproducer=medialab,pagebackref=true]{hyperref}
\hypersetup{colorlinks=true,urlcolor=blue,anchorcolor=blue,citecolor=blue,filecolor=blue,linkcolor=blue,menucolor=blue,pdfproducer=medialab}

\usepackage[utf8]{inputenc}
\usepackage[DIV=13]{typearea}
\usepackage{amsfonts}
\usepackage{amsmath,amssymb,amsthm}
\usepackage{bbm}
\usepackage{soul}
\usepackage{bm}
\usepackage{color}
\usepackage{slashed}
\usepackage[most]{tcolorbox}
\usepackage[toc,page]{appendix}
\usepackage{graphicx}
\usepackage{placeins}
\usepackage{caption}
\usepackage{xcolor}
\usepackage{subfigure}
\usepackage{booktabs}
\usepackage{microtype} 
\usepackage{comment}
\usepackage{physics}
\usepackage{float}
\usepackage{bbold}
\usepackage{ragged2e}
\usepackage{dcolumn}
\usepackage{orcidlink}
\usepackage{cite}
\usepackage[normalem]{ulem}

\DeclareGraphicsExtensions{.pdf,.png,.jpg}

\newcommand{\email}[1]{\href{mailto:#1}{\tt #1}}
\definecolor{airforceblue}{rgb}{0.36, 0.54, 0.66}
\definecolor{steelblue}{rgb}{0.27, 0.51, 0.71}
\definecolor{amber}{rgb}{1.0, 0.49, 0.0}
\definecolor{darkgreen}{rgb}{0.0, 0.5, 0.0}
\definecolor{amber}{rgb}{1.0, 0.49, 0.0}

\makeatletter \@addtoreset{equation}{section} \makeatother

\newcommand{\si}{\vec{i}}
\newcommand{\sj}{\vec{j}}
\newcommand{\sk}{\vec{k}}
\newcommand{\skp}{\vec{p}}
\newcommand{\sn}{\vec{n}}

\newcommand{\sy}{\vec{y}}

\newcommand{\ai}{[\vec{i}]}
\newcommand{\aj}{[\vec{j}]}
\newcommand{\ak}{[\vec{k}]}

\let\vec\mathbf

\begin{document}
\flushbottom
\begin{titlepage}
\vspace*{-1cm}
\hfill IPPP/26/71 \\
\vspace{1cm}

\begin{centering}
{\LARGE{\bf Vector Dark Matter  \\ \vspace{0.4cm} from $d\geq 2$ Extra Dimensions}} 

\vspace{1cm}

{ A.~de~Giorgi$^{\rm a\,}$\footnote{\email{arturo.de-giorgi@durham.ac.uk}}~\orcidlink{0000-0002-9260-5466}
and S.~Vogl$^{\rm b\,}$\footnote{\email{stefan.vogl@physik.uni-freiburg.de}}~\orcidlink{0000-0002-3005-9279}}

\vspace{0.8cm}

{\em $^{\rm{a}}$Institute for Particle Physics Phenomenology, Department of Physics, Durham University, Durham DH1 3LE, U.K.} 
\\
\vspace{0.15 cm}
{\em $^{\rm{b}}$University of Freiburg, Institute of Physics\\
Hermann-Herder-Str. 3, 79104 Freiburg,
Germany}
\vspace*{0.8cm}

\end{centering}
\vspace*{0.3cm}
 
\noindent

\textbf{Abstract}: In this work, we propose a new natural dark matter candidate: Kaluza-Klein towers of vectors from $d\geq 2$ extra-dimensional models.  
Their naturalness follows from two key features: (i) they arise unavoidably from the Kaluza-Klein decomposition of the higher-dimensional metric, and (ii) if the Standard Model is localised on a brane, they do not couple directly to it and interact only gravitationally through the other components of the higher-dimensional metric.
Remarkably, the rich particle content comes with as few as a single extra parameter. We first explain the general idea for $d\geq 2$, then focus on the $d=2$ case as our benchmark scenario. We derive the field content and interactions among the model particle content, and show how the geometry of the extra dimension unitarises the relevant processes for cosmology. We then compute the multi-component relic abundances of the vectors from freeze-in and discuss the available parameter space. 

\vfill
\end{titlepage}
\setcounter{footnote}{0}

\tableofcontents
\section{Introduction}

Extra-dimensional models have long been considered as an explanation for dark matter~(DM), both as a portal connecting the visible and dark sectors~\cite{Lee:2013bua,Lee:2014caa,Rueter:2017nbk,Folgado:2019gie,Folgado:2019sgz,Bernal:2020fvw, deGiorgi:2021xvm,deGiorgi:2022yha,Voronchikhin:2023znz,Voronchikhin:2023qig,Chivukula:2024nzt,Lee:2024wes,Chivukula:2025pmk,Donini:2025cpl,Donini:2025qrf} and as the origin of the DM candidate \cite{Servant:2002aq,Cheng:2002ej,Hooper:2007qk,Gonzalo:2022jac,Anchordoqui:2022svl,Obied:2023clp,Koutroulis:2024wjl} itself.
However, despite the considerable interest in these models, there are still intriguing possibilities that have not been considered so far. Traditionally, extra-dimensional frameworks have been shaped by theoretical biases towards constructions that solve the hierarchy problem. This prejudice has limited the range of considered models and the parameter choices that control the phenomenology. Relaxing this assumption leads to intriguing alternatives with distinct phenomenology \cite{Montero:2022prj,Law-Smith:2023czn,deGiorgi:2024elx,Gross:2024zdh,Im:2024kuw,Hardy:2025ajb,deGiorgi:2026isr,Clarke:2026cdl}. 

At the same time, the phenomenology of theories with several extra dimensions remains comparatively underexplored, despite being well motivated by ultraviolet constructions such as string theory, in part due to the rapidly increasing complexity of their Kaluza-Klein sectors. Indeed, theories with multiple extra dimensions contain a rich spectrum of purely gravitationally interacting states, arising from the Kaluza-Klein~(KK) decomposition of the extra-dimensional metric. These include spin-2, spin-1, and spin-0 states. It is worth pointing out that this is true already before any matter content is introduced. The spin-1 states, the so-called ``graviphotons'', do not couple directly to matter fields localized on a brane; see e.g. Refs.~\cite{Han:1998sg,Giudice:1998ck}. The suppressed couplings make them harder to detect than the massive graviton counterparts, and thus their phenomenology has largely been overlooked.

In this work, we take the opposite perspective and exploit this feature. We consider models with two or more large extra dimensions where the Standard Model~(SM) is localised on a brane and argue that such vector components are natural dark matter candidates. 
In fact, they (a) source gravitational fields like any other particle, (b) are naturally dark due to their exclusively gravitational interactions with the SM, and (c) potentially remain stable over cosmological timescales, as their decays are only permitted through higher-order Planck-suppressed operators.

The latter point is particularly inherent to the lightest graviphoton. As it does not couple directly to the SM, the only conceivable two-body decays are into the zero-modes of the KK tower. However,  a massive vector cannot decay into two massless vectors, as forbidden by the Landau-Yang theorem, nor into two massless gravitons or a pair of massless scalars; see e.g. Refs.~\cite{Strumia:2025dfn} and~\cite{Pleitez:2015cpa}. Consequently, even if some modes of the tower are unstable, the lightest modes of the graviphotons cannot decay into any of the gravitational particle content. This forces them to decay into higher multiplicity SM final states via off-shell KK-particles. Such processes are only possible at the price of additional Planck-mass suppressions, making them naturally exceptionally stable.

While the general construction holds for any number $d\geq2$ of extra dimensions, the field content and the structure of the interactions become increasingly involved as $d$ grows. We therefore present the general setup for arbitrary $d$, and then specialise to $d=2$ as our benchmark scenario. We argue that the qualitative features of this benchmark, in particular the multi-component dark matter structure and the freeze-in production mechanism, carry over to higher $d$. For the $d=2$ case, we derive the full field content and interactions of the model, and show how the geometry of the extra dimension unitarises the scattering processes relevant for cosmology. We then compute the yield of each component of the Kaluza-Klein tower produced via freeze-in, and use these results to map out the viable parameter space for the model to account for the observed dark matter abundance. 
In this context, it becomes relevant to determine whether KK modes can decay into each other. As will be detailed later, in a flat extra-dimensional model with only gravity propagating in the bulk, extra-dimensional translational invariance, jointly with the phase space factor, ensures stability of the KK modes. 
We label such a case as the ``KK-conserving'' scenario. Translational invariance is broken by the brane where the SM lives, and thus it is the minimal source of KK-number violation. Then, the KK-vector tower will be largely stable, and one has to deal with a multi-component dark matter sector that is dominated by heavy modes. In models with additional sources of KK-violation, this need not hold, and decays between KK-tower states become relevant. Using a simple parametrization of such a violation, we identify a benchmark model for this scenario and investigate its phenomenology. We would like to caution, however,  that different sources of KK-breaking may well enable alternative decay patterns. Therefore, this should be treated as a motivated benchmark model and not as the universal result for dark decays of the KK-vectors.

\bigskip
The remainder of this paper is organised as follows. In Sec.~\ref{sec:model} we introduce the minimal $D=4+d$ dimensional model, presenting the gravitational particle content and their couplings among themselves and with the SM. In Sec.~\ref{sec:phenomenology} we discuss the cross-sections needed for the cosmological computation of the model. We further show how an intricate series of cancellations ensured by the extra-dimensional geometry unitarises them in the high-energy limit. In Sec.~\ref{sec:early-universe} we compute the dark matter yield from freeze-in, separately treating the relativistic, non-relativistic, and intermediate regimes, and present our results and the resulting constraints on the parameter space. Finally, we draw our conclusion and outline future work directions in Sec.~\ref{sec:conclusions}.

\section{The minimal $D=4+d$ dimensional model}
\label{sec:model}

Our starting point for the following discussion is the Einstein-Hilbert action in $D=4+d$ dimensions~\cite{Antoniadis:1990ew,Arkani-Hamed:1998jmv,Antoniadis:1998ig}
\begin{equation}
    S_D=\frac{M_D^{D-2}}{2}\int d^4x\int_{\mathcal{M}_d} d^dy\,\sqrt{|G|}\mathcal{R}\,,
\end{equation}
where $G$ is the metric determinant, $\mathcal{R}$ is the Ricci scalar and $M_D$ denotes the $D$ dimensional Planck mass. We assume that the extra $d$-dimensions are compact. For simplicity, we will take them to be compactified on a $\mathcal{M}_d=(S_1)^d/\mathbb{Z}_2$ orbifold with universal radius $R$. 
Notation-wise, capital Latin letters $\{A,B,\dots\}$ are used to indicate coordinates over the $D$-dimensional manifold, $x^A$; extra-dimensional coordinates are denoted by $\{y^a\}_{a=1}^d$, while we reserve the usual $\{x^\mu\}_{\mu=0}^3$ notation for the four-dimensional Minkowski ones. 
The extra-dimensional Planck mass is related to the 4D one by
\begin{equation}
    \bar{M}_{P}^2\equiv M_D^{D-2} V_d= M_D^{d+2} (2\pi R)^d\,,
\end{equation}
where $\bar{M}_P\approx 2.4\times 10^{18}$~GeV is the reduced Planck mass and $V_d=(2\pi R)^d$ is the volume of the extra-dimensional space.
The SM Lagrangian $\mathcal{L}_\text{SM}$ is taken to live on one of the orbifold fixed points 
\begin{equation}
    S_\text{SM}= \int d^4x\int_{\mathcal{M}_d} d^dy\,\delta(\sy-\sy_\star)~\sqrt{|G|}\mathcal{L}_\text{SM}\,,
\end{equation}
Given the flatness of the space, all orbifold fixed points are equivalent. For the sake of concreteness, in the case of interest $d=2$,  we pick $\sy_\star=(\pi R,\pi R)$.

\subsection{Field Content}
\label{sec:fields}
The static background metric is a $D$-dimensional Minkowski space with, in the mostly minus convention,
\begin{equation}
    ds^2=\eta_{\mu\nu}dx^\mu dx^\nu+\eta_{ab}dy^a dy^b\equiv(dx^0)^2-\sum\limits_{i=1}^3(dx^i)^2-\sum\limits_{a=1}^d(dy^a)^2\,.
\end{equation}
The gravitational particle content is obtained by perturbing the background metric
\begin{align}
    \label{eq:weak-exp}&G_{AB}= \eta_{AB}+\kappa H_{AB}(\vec{x},\vec{y})\,, &\kappa\equiv \frac{2}{M_D^{(D-2)/2}}\,.
\end{align}
 The components of $H_{\mu\nu}$, $H_{\mu a}$ and $H_{ab}$ correspond to tensorial, vectorial and scalar excitations in the 4D EFT.
Their transformation properties under $\vec{y}\to-\vec{y}$ are constrained by the orbifold symmetry.
In general, a coordinate transformation $x^A\to \Tilde{x}^A(x^B)$ the metric transforms as
\begin{equation}
    \tilde G_{AB}(\tilde x)=\frac{\partial x^M}{\partial \tilde x^A}\frac{\partial x^N}{\partial \tilde x^B} G_{MN}(x)\,.
\end{equation}
It implies that under the reflection of $\vec{y}$-coordinates, $\vec{y}\to -\vec{y}$ one has
\begin{align}
    &\begin{cases}
        \tilde G_{\mu\nu}= G_{\mu\nu}\,,\\
        \tilde G_{\mu a}= -G_{\mu a}\,,\\
        \tilde G_{ab}= G_{ab}\,,
    \end{cases}
    &&\Rightarrow
    &&\begin{cases}
        \tilde H_{\mu\nu}(x,\vec{y})= H_{\mu\nu}(x,-\vec{y})\,,\\
        \tilde H_{\mu a}(x,\vec{y})= -H_{\mu a}(x,-\vec{y})\,,\\
        \tilde H_{ab}(x,\vec{y})= H_{ab}(x,-\vec{y})\,.
    \end{cases}
\end{align}
Thus, the orbifold naturally assigns negative parity to the vectorial component and positive parity to the scalar and tensorial ones.

\bigskip
The particle content in the 4D EFT can be obtained from $H_{AB}$ by integrating out the extra dimensions. This can be done by first expanding the extra-dimensional fields in terms of Kaluza-Klein modes
\begin{equation}
    \label{eq:KK-decomposition}H_{AB}(\vec{x},\vec{y})=\frac{1}{\sqrt{V_d}}\sum\limits_{\vec{n}} H^{(\vec{n})}_{AB}(\vec{x})\psi_{\vec{n}}(\vec{y})\,,
\end{equation}
where $\{\psi_{\vec{n}}(\vec{y})\}_\sn$, which are typically referred to as ``wave-functions''~(WFs), constitute a set of orthonormal functions for the extra-dimensional manifold.
Notation-wise, from now on we will employ Latin indices from the letter ``i'' onward as $\{\si,\sj,\dots \sn\}$ to indicate the components of the KK-tower, in contrast to $\{a,b,c,\dots\}$ which are reserved to label extra-dimensional components.

The equations of motion of the WFs and their normalisation condition read
\begin{align}
    &\left(\sum\limits_{a=1}^d \partial_a^2 + m_{\Vec{n}}^2\right)\psi_{\Vec{n}}(\sy) =0\,, &\frac{1}{(2\pi R)^d}\int d^dy\ \psi_{\Vec{i}}(\vec{y})\psi_{\vec{j}}(\vec{y}) =  \delta_{\vec{i}\vec{j}} \,.
\end{align}
Depending on the parity of the associated field, there are two possibilities, which we label with only $+$ or only $-$.
The 0-mode has a non-trivial solution only for the positive parity and yields
\begin{equation}
    \psi_{\vec{0}}(\vec{y})=1\,.
\end{equation}
with associated $m_{\vec{0}}=0$.
The solution for the massive cases yields
\begin{align}
   &\psi^+_{\Vec{n}}(\Vec{y})=\sqrt{2}\cos\left(\frac{\vec{n}}{R}\cdot\vec{y}\right)\,, &&\psi^-_{\Vec{n}}(\Vec{y})=\sqrt{2}\sin\left(\frac{\vec{n}}{R}\cdot\vec{y}\right)\,.
\end{align}
and the mass spectrum is given by
\begin{equation}
    m_{\vec{n}}^2=\frac{|\vec{n}|^2}{R^2}\,,
\end{equation}
where $\vec{n}=(n_1,\dots n_d)$ is a vector composed of integers.

\begin{figure}
    \centering
    \includegraphics[width=0.7\textwidth]{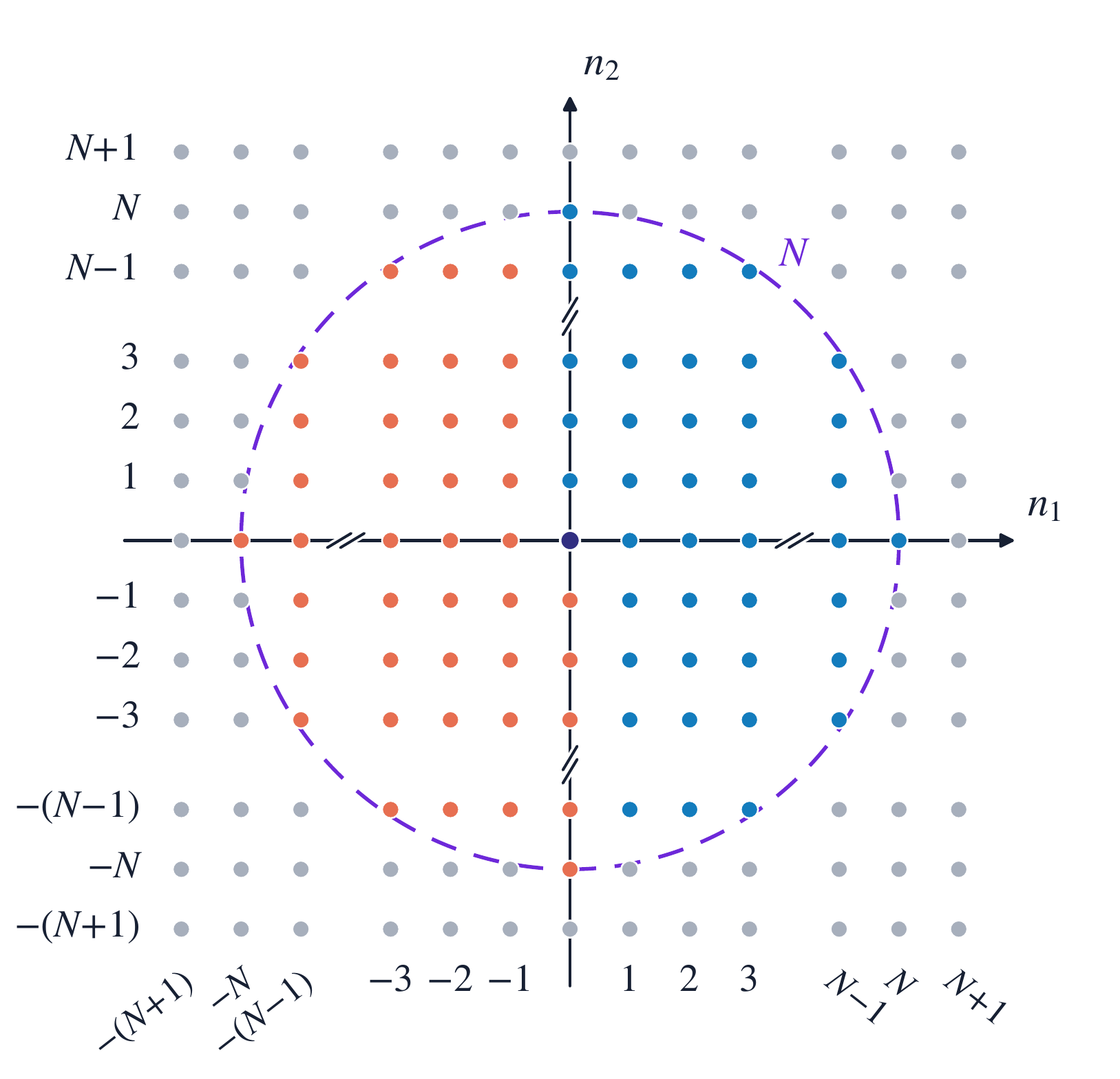}
    \caption{Grid of non-redundant KK-vectors allowed in our convention (blue), redundant (red), and parity-dependent (purple). The dashed purple circle highlights the vectors allowed by the cut-off of the tower and the excluded ones in gray, defined such that $|\vec{n}|^2=n_1^2+n_2^2\leq N^2$.}
    \label{fig:KK-vectors}
\end{figure}
The values of $\vec{n}$ that are allowed must not be redundant. The cosine and sine functions come from combining the Fourier modes stemming from the extra-dimensional Poisson equation. 
In a sense, if $\vec{n}$ is allowed, then the orbifold parity identifies it with $-\vec{n}$, so it does not need to be counted in the form we wrote the solutions.
We choose the convention for which 
\begin{equation}
    (n_1,n_2)\in \{(\mathbb{N},\mathbb{Z}) \cup (0,\mathbb{N}_0)\}\,,
\end{equation}
where $\mathbb{N}$ is the set of natural numbers and $\mathbb{N}_0$ also includes the zero. The grid of non-redundant KK vectors allowed is shown in Fig.~\ref{fig:KK-vectors}. 

The sum of Eq.~\eqref{eq:KK-decomposition} in principle scans over an infinite number of 2D vectors. However, as $|\vec{n}|$ grows, so does the mass of the associated state, which must not exceed the cut-off of the theory. Therefore, the tower of states needs to be truncated.
If we select some cut-off $\Lambda$, then the largest KK vectors allowed must satisfy $m_\sn<\Lambda$. In the absence of further new physics, it comes naturally to associate the cut-off with the six-dimensional Planck mass $\Lambda = M_6$. We will denote the largest KK vector for which the expansion is acceptable by $N\equiv |\vec{n}_\text{max}|$, for which $m_N= M_6$. This is illustrated in Fig.~\ref{fig:KK-vectors} by a purple dashed line.

\bigskip
The choice of coordinates in the action amounts to gauge freedom for $H_{AB}$, which needs to be fixed to obtain the physical particle content. 
The degrees of freedom~(DOF) of a symmetric field $H_{AB}$ in $D$ dimensions are given by $D(D+1)/2$. They can first be reduced by imposing the so-called ``harmonic gauge''
\begin{equation}
    H_{AB}^{,B}=\frac{1}{2}H_{,A}\,,
\end{equation}
which removes $D$ DOF. However, this still leaves a gauge residual, which amounts to the freedom to redefine the coordinates $x^A\to x^A+\epsilon^A$ as long as
\begin{equation}
    \Box \epsilon^A=0\,,
\end{equation}
which removes another $D$ DOF. Their total number after gauge-fixing is therefore
\begin{equation}
    \# H_{AB}=\frac{D(D+1)}{2}-2D=\frac{D(D-3)}{2}=\frac{(1+d)(4+d)}{2}\,.
\end{equation}
This leads to $\# H_{AB}=5$ in  $d=1$ and $\# H_{AB}=9$ in $d=2$ and so on.

Let us now analyse the type of DOF produced in the 4D EFT. For each massive KK mode, the purely four-dimensional part of the metric contributes a tower of massive spin-2 field: the KK-graviton, with 5 DOF. In $d=1$, this saturates the DOF, and it is indeed the only massive set of states. In $d\geq 2$, also vectorial and scalar modes become relevant. Indeed, including the transverse and traceless conditions (see later Eq.~\eqref{eq:transverse-conditions} and related discussion), restricts the number of independent massive fields to include $\#V$ massive vectors and $\#S$ scalars given by
\begin{align}
    &\#V=d-1\,, &&\#S=\frac{d^2-d}{2}\,,
\end{align}

The counting applied to the massive states does not apply to the massless ones, and the conditions mentioned above and presented later in Eq.~\eqref{eq:transverse-conditions} do not apply to the massless modes. Due to the odd nature of the vectors, there are no zero modes, while the number of scalar  zero modes is given by 
\begin{equation}
    \#S_0=\frac{d(d+1)}{2}\,.
\end{equation}

In $d=1$, there is a single massless scalar, typically dubbed as the ``radion''. For $d =2$ there are three massless scalars: the radion, which corresponds to the trace of the perturbation, the diagonal mode $D_0$ and the corner one $C_0$
\begin{align}
    \label{eq:extra-scalars}&H^a_a=\phi_0\,, &&H_{12}=H_{21}=\frac{C_0}{\sqrt{2}}\,, &&H_{11}=-H_{22}=\frac{D_0}{\sqrt{2}}\,.
\end{align}

\bigskip
A possible choice to implement the above counting of DOF within the field parametrisation is the so-called ``unitary'' gauge~\cite{Giudice:1998ck}, under which the metric perturbation can be decomposed in terms of the set of fields $\{h_{\mu\nu},V_{\mu i},S_{ij},\phi\}$. In the case of interest $d=2$, the following parametrization allows one to get the physical particle content and a canonically normalised Lagrangian
\begin{align}
    &\label{eq:un1} H^a_a(\vec{x},\vec{y})=\phi^{(\vec{0})}+\sum\limits_{\vec{n}\neq \vec{0}}\sqrt{\frac{3}{4}} \phi^{(\vec{n})}\,,\\
    &\label{eq:un2} H_{ab}(\vec{x},\vec{y})=\frac{\eta_{ab}}{2}\phi^{(\vec{0})}+\sum\limits_{\vec{n}\neq \vec{0}}\left[S_{ab}^{(\vec{n})}+\sqrt{\frac{3}{4}}\left(\eta_{ab}+\frac{m_{\vec{n},a}m_{\vec{n},b}}{m_{\vec{n}}^2}\right)\phi^{(\vec{n})}\right]\,,\\
    &\label{eq:un3} H_{\mu a}(\vec{x},\vec{y})=\sum\limits_{\vec{n}\neq \vec{0}}\frac{V_{\mu a}^{(\vec{n})}}{\sqrt{2}}\,,\\
    &\label{eq:un4} H_{\mu\nu}(\vec{x},\vec{y})=\left(h_{\mu\nu}^{(\vec{0})}-\frac{1}{2}\eta_{\mu\nu}\phi^{(\vec{0})}\right)+\sum\limits_{\vec{n}\neq \vec{0}}\left[h_{\mu\nu}^{(\vec{n})}-\sqrt{\frac{1}{12}}\left(\eta_{\mu\nu}+\frac{\partial_\mu\partial_\nu}{m_{\vec{n}}^2}\right)\phi^{(\vec{n})}\right]\,,
\end{align}
where $ m_{\vec{n},a}$ are vector components of the total mass squared of the particle such that
\begin{equation}
    m_{\vec{n}}^2\equiv m_{\vec{n},a} m_{\vec{n},a}=m_{\vec{n}}^a m_{\vec{n}}^a=-m_{\vec{n},a} m_{\vec{n}}^a\,,
\end{equation}
and for the sake of readability, we left implicit the WF dependence, and each field $\Psi\in\{h_{\mu\nu},V_{\mu a},S_{ab},\phi\}$ appearing in the above equations should be understood as
\begin{equation}
    \Psi^{(\vec{n})}\to \Psi^{(\vec{n})}(\vec{x})\left(\frac{\psi^\pm_{\vec{n}}(\vec{y})}{\sqrt{V_d}}\right)\,.
\end{equation}
Finally, the fields $V_{\mu a}$ and $S_{ab}$ satisfy the transversality and traceless conditions
\begin{align}
    &\label{eq:transverse-conditions}m_\sn^a V_{\mu a}^{(\sn)}=0\,, && S^a_{\sn,a}=0 &&m_\sn^a S_{ab}^{(\sn)}=0\,.
\end{align}

The vectorial components $V_{\mu a}$ still carry non-trivial extra-dimensional dependence, and only one combination corresponds to the physical fields.
In $d=2$, it can be obtained by employing the expansion
\begin{equation}
    V_{\mu a}(\vec{x},\vec{y})=\frac{1}{(2\pi R)}\sum\limits_{\vec{n}}V_{\mu a}^{(\sn)}(\vec{x})\psi^-_{\sn}(\vec{y})\equiv \frac{1}{(2\pi R)}\sum\limits_{\vec{n}} V_{\mu}^{(\vec{n})}(\vec{x})\chi_{\sn,a}(\vec{y})\,,
\end{equation}
where $V^\mu_\sn$ is now the physical field.
The orthogonality and normalisation conditions fix the vector $\chi_{n}^a(\vec{y})$ to be
\begin{align}
    &\chi_{\sn}^a(\vec{y})=\psi_\sn^-(\vec{y})\,\frac{m_{\vec{n},b}\epsilon^{ba}}{m_{\vec{n}}}=\psi_{\vec{n}}^-(\vec{y})\begin{pmatrix}
        \frac{m_{\vec{n}}^{(2)}}{m_{\vec{n}}}\,, -\frac{m_{\vec{n}}^{(1)}}{m_{\vec{n}}}
    \end{pmatrix}\,, 
\end{align}
where $\epsilon^{ab}$ is the totally anti-symmetric tensor defined such that $\epsilon^{12}=\epsilon_{12}=1$.


\subsection{Couplings}
\label{sec:couplings}
The lengthy procedure detailed in the previous section fixes the gravitational sector of the theory, through which interactions among all the different modes can be obtained, as well as the coupling interactions with matter. Let us begin analysing the latter.

The coupling with matter at leading order in the $\kappa$ expansion of the metric reads
\begin{equation}
    \label{eq:couplings}\mathcal{L}\supset -\left.\frac{\kappa}{2}H_{\mu\nu}T^{\mu\nu}\right|_{\sy=\sy_\star}=-\left.\frac{\kappa}{2}\left[\left(h_{\mu\nu}^{(\vec{0})}+\sum\limits_{\vec{n}\neq \vec{0}}h_{\mu\nu}^{(\vec{n})}\right)T^{\mu\nu}-\frac{1}{2}\left(\phi^{(\vec{0})}+\sum\limits_{\vec{n}\neq \vec{0}}\sqrt{\frac{1}{3}}\phi^{(\vec{n})}\right)T\right]\right|_{\sy=\sy_\star}\,,
\end{equation}
where $T^{\mu\nu}$ is the energy-momentum tensor of the SM and $T=\eta_{\mu\nu}T^{\mu\nu}$ its trace. The effective coupling of each 4D mode is then proportional to
\begin{equation}
    \frac{\kappa}{2}\left(\frac{\psi_\sn(\sy_\star)}{\sqrt{V_d}}\right)=\frac{1}{\bar{M}_P}\times\begin{cases}
        1 & \sn=\vec{0}\,,\\
        \sqrt{2} &\sn \neq \vec{0}\,.
    \end{cases}
\end{equation}
As can be seen, out of the spectrum discussed in the previous section, only the towers of massive gravitons and radions couple directly to the SM. A tower of massive vectors and two scalar zero-modes of Eq.~\eqref{eq:extra-scalars} couple directly only to the gravitational sector itself at order $\mathcal{O}(\bar{M}_P^{-1})$. 

\bigskip
The interaction structure involving the different towers of KK states is far more complicated, but can be found by systematically expanding the Lagrangian as detailed in the previous section.
The strength of the interactions is dictated by overlap integrals over the WFs stemming from dimensional reduction of the Lagrangian. Employing the EOMs, all extra-dimensional derivatives can be converted to mass terms, and ultimately, all interactions of interest for this work depend on the coefficients
\begin{align}
    &\label{eq:a-coefficients}a_{\vec{i}\vec{j}\vec{k}}\equiv \dfrac{1}{(2\pi R)^2}\int d^2y\,\psi_{\vec{i}}^+ \psi_{\vec{j}}^+ \psi_{\vec{k}}^+\,, &a_{\vec{i}[\vec{j}][\vec{k}]}\equiv \dfrac{1}{(2\pi R)^2}\int d^2y\,\psi_{\vec{i}}^+ \psi_{\vec{j}}^- \psi_{\vec{k}}^-\,,
\end{align}
where the square brackets are used to identify overlap integrals of odd-parity WFs. Other combinations of WFs are trivially zero due to parity.
The two types of coefficients are related to each other, as
\begin{align}
    0&=\int d^2y\,\partial_a\left(\partial_a(\psi_\si^+)\psi_\sj^+\psi_\sk^+\right)=-m_\si^2a_{\si\sj\sk}+m_{\si\sj}a_{\ai\aj \sk}+m_{\si\sk}a_{\ai \sj\ak}\,,  &m_{\vec{i}\vec{j}}\equiv m_{\vec{i}}^am_{\vec{j}}^a\,,
\end{align}
and thus
\begin{equation}
    m_\si^2a_{\si\sj\sk}=m_{\si\sj}a_{\ai\aj \sk}+m_{\si\sk}a_{\ai \sj\ak}\,.
\end{equation}
By taking appropriate combinations, one can then relate mixed coefficients of Eq.~\eqref{eq:a-coefficients}
\begin{align}
\label{eq:master-a-equation}
    a_{\si\aj\ak}m_{\sj\sk}=\frac{1}{2}a_{\si\sj\sk}\left(m_\sj^2+m_\sk^2-m_\si^2\right)\,.
\end{align}
Ultimately, this allows us to work solely with $a_{\vec{i}\vec{j}\vec{k}}$.
Their structure is non-trivial: due to KK number conservation, these overlap integrals are non-zero only if any combination of sum or difference of the KK-vectors $\vec{i},\vec{j},\vec{k}$ is zero; a summary of their values is
\begin{equation}
\label{eq:CI-KKselection}
    a_{\vec{i}\vec{j}\vec{k}}=\delta_{\vec{i}\pm\vec{j}\pm\vec{k},\vec{0}}~\times\begin{cases}
        1 & \si=\sj=\sk=\vec{0}\,, \\
        1 & \si=\vec{0}\,,~\sj,\sk\neq 0 \\
        \frac{1}{\sqrt{2}} & \si,\sj,\sk\neq 0\,, \\
    \end{cases}\,.
\end{equation}


\section{Decays, Cross-Sections and Unitarisation}
\label{sec:phenomenology}

Equipped with the interactions of the fields, we now turn to the study of the decay and scattering processes necessary to study the cosmology of the model.

Before entering details, a comment about the massless scalars of the theory described in Eq.~\eqref{eq:extra-scalars} is in order. The presence of gravitationally coupled massless particles is phenomenologically problematic, as they would mediate long-range fifth forces. We therefore assume a mechanism that generates a mass for the scalar mode, e.g. the Goldberger-Wise type~\cite{Goldberger:1999uk} is at work. For simplicity, and to extract the relevant features of the dark matter phenomenology, we work with the mass spectrum of the unstabilised theory, as this preserves the delicate cancellation structure of the amplitudes (see Sec.~\ref{sec:cross-section} and App.~\ref{app:high-energy} for further details). The precise implementation of the stabilisation mechanism and its effect on the spectrum is left for future work.

\bigskip
In order to simplify the notation, we will occasionally write $i$ rather than $\vec{i}$, making sure the difference is clear from the context. For example, $V_i\equiv V_\si$, while $m_1(i+j)/T\equiv m_1(|\si|+|\sj|)/T$. For the calculations, it will also be useful to adopt a polar notation for the direction in KK vector space such that 
\begin{align}
    m_{\vec{i}}^a=|m_\si|\begin{pmatrix}
        \cos(\theta_\si)\\
        \sin(\theta_\si)
    \end{pmatrix}^a\,,
\end{align}
where $|m_\si|=m_1 |\si|$ and $\theta_i\in(-\pi/2,\pi/2]$. Given that the choice of KK vectors exemplified in Fig.~\ref{fig:KK-vectors} is arbitrary, the amplitudes can only depend on rotation-invariant quantities. In the case of interest they can be identified with $|\si|,|\sj|$, $(\si\cdot\sj)$, and $|\si\cross\sj|$ or, equivalently, $|\si|,|\sj|$ and the angle difference among them, $\Delta\equiv \theta_i-\theta_j$.

\subsection{Interaction Vertices and Computational Techniques}

To derive the relevant Feynman rules and compute observables necessary for the analysis, we implement the procedure detailed in Sec.~\ref{sec:fields}, expanding the Lagrangian at $\mathcal{O}(\kappa)$ as outlined in Eq.~\eqref{eq:weak-exp}. We then extract the physical degrees of freedom, impose the unitary gauge, expand the KK towers, and keep the relevant interactions.
The derivation of the interaction Lagrangian has been implemented via \texttt{Mathematica}~\cite{Mathematica} jointly with \texttt{xAct}~\cite{xact}. The latter is a collection of Mathematica packages for tensor computer algebra, particularly suited to the symbolic manipulation of tensorial expressions arising in differential geometry and GR.
It is worth noticing that, as a result of the expansion: 
\begin{itemize}
    \item[(i.)] The vertices $h_0 h_i V_i$, $\phi_0 h_i V_i$, $h_0 \phi_i V_i$ and $\phi_0 \phi_i V_i$ vanish at the Lagrangian level.
    \item[(ii.)]  Instead, the vertices $h_j h_i V_i$, $\phi_j h_i V_i$, $h_j \phi_i V_i$ and $\phi_j \phi_i V_i$, do appear but vanish once KK-conservation is imposed. 
\end{itemize}
Finally, the Feynman rules of the remaining non-vanishing interactions were derived via \texttt{FeynRules}~\cite{Alloul:2013bka}.

\bigskip
The symbolic calculations of the amplitudes were performed employing \texttt{Feyncalc}~\cite{Shtabovenko:2023idz}.
Given the involved structure of Feynman rules, the symbolic computation of scattering amplitudes involving spin-1 and spin-2 massive final states turns out to be computationally demanding. 
We therefore choose to compute all scattering amplitudes employing the helicity method. This has a double advantage. First, it makes the computation extremely fast compared to the ``brute-force'' approach of spin-averaging techniques directly over the squared amplitude. Secondly, it allows for a clean understanding of the cancellation between Feynman diagrams.
In this approach, each amplitude is then computed in terms of helicities of the final states, e.g., for two-graviton final states with helicities $\lambda_{1,2}$
\begin{equation}
    i\mathcal{M}_{\lambda_1,\lambda_2}\equiv i\mathcal{M}_{\mu\nu\alpha\beta}\times \epsilon_{\lambda_{1}}^{\mu\nu}\epsilon_{\lambda_{2}}^{\alpha\beta}\,,
\end{equation}
where $\epsilon_{\lambda_{1,2}}^{\mu\nu}$ are the polarisation tensors of a massive spin-2 state. An analogous decomposition holds for spin-1 states.

The spin-2 polarization tensor can be decomposed as a combination of spin-1 polarization tensors \cite{Gleisberg:2003ue}
\begin{align}
       \label{pol-tens-2}&\epsilon_{0}^{\mu\nu}  = \frac{1}{\sqrt{6}}\left( \epsilon_{\pm 1}^{\mu}\epsilon_{\mp 1}^{\nu}+2 \epsilon_{0}^{\mu}\epsilon_{0}^{\nu}+ \epsilon_{\mp1}^{\mu}\epsilon_{\pm 1}^{\nu}\right) \ ,
       &&\epsilon_{\pm 1}^{\mu\nu}  = \frac{1}{\sqrt{2}}\left(\epsilon_{\pm 1}^{\mu}\epsilon_{0}^{\nu}+ \epsilon_{0}^{\mu}\epsilon_{\pm 1}^{\nu}\right)\ , 
       &&\epsilon_{\pm 2}^{\mu\nu}  = \epsilon_{\pm 1}^{\mu}\epsilon_{\pm 1}^{\nu} \,.
\end{align}
 An explicit form for the spin-1 polarization tensor of a massive particle moving in an arbitrary direction, i.e. a particle with mass $m$ and momentum $\Vec{p}=|\vec{p}|\left(\sin\theta\cos\phi,\sin\theta\sin\phi,\cos\theta \right)$,  is given by
\begin{align}
\label{pol-tens-1}
       \epsilon_{\pm 1}^{\mu} & = \frac{1}{\sqrt{2}}e^{\mp i \gamma}\left(0;\mp \cos\theta\cos\phi + i \sin\phi,\cos\theta\sin\phi -i \cos\phi,\pm \sin\theta \right) \ , \\
       \epsilon_{0}^{\mu} & =\frac{E}{m}\left(\sqrt{1-\frac{m^2}{E^2}};\sin\theta\cos\phi,\sin\theta\sin\phi,\cos\theta\right)\ .
\end{align}
For the processes of interest, the angles $\phi$ and $\gamma$ can be chosen to be zero without loss of generality.
The full unpolarised scattering matrix element can be computed by appropriately summing over all helicities. For example, given two final states with helicities $\lambda_{1,2}$, the unpolarised amplitude squared can be obtained via
\begin{equation}
    \label{eq:helicity-squared}|\mathcal{M}|^2=\sum\limits_{\lambda_1,\lambda_2}|\mathcal{M}_{\lambda_1,\lambda_2}|^2=\Tr{\mathcal{M}_{\lambda_1,\lambda_2}\mathcal{M}_{\lambda_1,\lambda_2}^\dagger}\,.
\end{equation}

\subsection{Decay}
\label{sec:decay}
The decay channels of the DM candidates play a central role. While gravitons and scalars decay mainly to SM fields via the interactions of Eq.~\eqref{eq:couplings}, the decay channels of the graviphotons are very sensitive to the model under consideration. As we will show, the flatness of the extra dimensions strongly affects the decay channels of the KK states via KK number conservation, particularly the graviphotons. In this section, we first discuss the consequences of this flat-space model, then examine how deviations from KK number conservation modify the phenomenology.

\subsubsection{KK Number Conservation}
In the framework under consideration,  decays of any KK mode  $x_{\vec{i}} \in \{\phi_{\vec{i}}, V_{\vec{i}}, h_{\vec{i}}\}$  into combinations of other KK modes,
\begin{equation}
    x_{\vec{i}} \to x_{\vec{i}_1}\, x_{\vec{i}_2}\, \cdots\, x_{\vec{i}_n}\,,
\end{equation}
are forbidden at tree-level. This follows from an interplay between the KK selection rule of Eq.~\eqref{eq:CI-KKselection} and the available phase space. Indeed, the KK selection rule requires KK number conservation at each vertex, for example
\begin{equation}
    \vec{i} = \vec{i}_1 + \vec{i}_2 + \cdots + \vec{i}_n\,.
\end{equation}
Together with the mass relation $m_{\vec{i}} = |\vec{i}|\, m_1$, the triangle inequality gives
\begin{equation}
    \label{eq:mass-inequality}m_{\vec{i}} = |\vec{i}_1 + \vec{i}_2 + \cdots + \vec{i}_n|\, m_1
    \leq \left(|\vec{i}_1| + |\vec{i}_2| + \cdots + |\vec{i}_n|\right) m_1
    = \sum_{k=1}^{n} m_{\vec{i}_k}\,.
\end{equation}
Therefore, the mass of the decaying particle never exceeds the sum of the decay products, rendering the decay kinematically forbidden regardless of the specific mode or final state multiplicity $n$.

\begin{figure}
    \centering
    \includegraphics[width=0.5\linewidth]{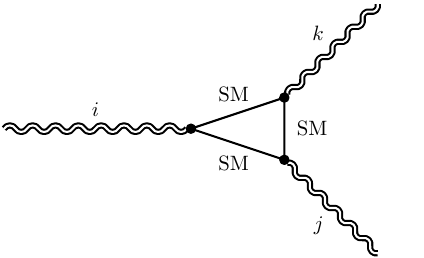}
    \caption{Representative diagram for the loop-induced, KK-number-violating decay of $h_i$ into lighter gravitons $h_j h_k$. }
    \label{fig:loop-decay-vectors}
\end{figure}
Such conclusions hold only at tree level. The KK number conservation of Eq.~\eqref{eq:CI-KKselection}  follows from translation invariance along the fifth dimension, which is broken by the brane on which the SM matter fields are localised. Once brane-localised matter or loop effects are included, KK-number-violating processes become possible. 
An example of such a process can be seen in Fig.~\ref{fig:loop-decay-vectors}. The brane-localised SM matter running in the loop breaks the fifth-dimensional translation invariance responsible for the tree-level selection rule of Eq.~\eqref{eq:CI-KKselection}, thus allowing the decay of a graviton $h_\si$ into an arbitrary combination of gravitons $h_\sj h_\sk$.
However, such contributions are 
suppressed by additional powers of $\bar{M}_P$ and loop factors. Therefore, we expect them to be negligible for the observables studied in this work. 

\begin{figure}
    \centering
    \includegraphics[width=\linewidth]{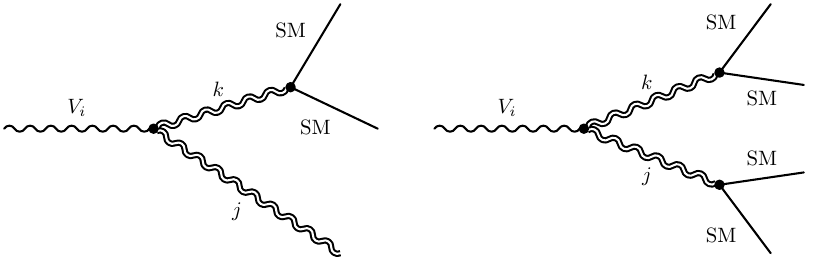}
    \caption{Representative set of Feynman diagrams responsible for the tree-level decay of vectors into final states which include SM fields.}
    \label{fig:decay-vectors}
\end{figure}

As discussed in Sec.~\ref{sec:couplings}, the graviphotons $V_{\vec{i}}$, do not interact directly with SM fields, and thus their decay rate is absent at lowest order in $\bar{M}_{Pl}$. However, via the brane, heavier vector modes can decay into a massive graviton and a SM pair from a massive off-shell mode, as shown in the left diagram of Fig.~\ref{fig:decay-vectors}.  This three-body decay is the leading decay channel for heavy graviphotons; however, it does not apply to the
lightest graviphoton. As the vertex $V_i h_jh_0$ does not exist, and the vertex $h_ih_jh_0$ only exists for $i=j$, the decay of the lightest vector into two SM fields and a single massless graviton is not possible. Thus the only viable channel is into two pairs of SM final states mediated by off-shell KK gravitons or graviscalars, as exemplified in the right diagram of Fig.~\ref{fig:decay-vectors}.  

In practice, the only phenomenologically relevant decays to set up constraints in this setup are those of KK gravitons $h_{\vec{i}}$ and KK scalars $\phi_{\vec{i}}$ into SM fields, which we include in the analysis whenever necessary. 

\subsubsection{KK Number Violation}
Before moving to the production part, we first comment on the impact of deviations from the KK number-conserving limit.
The previous conclusions were strongly dependent on the absolute stability of KK modes. This was ensured by the interplay between the kinematically available channels and the non-zero three-point overlap structure due to KK-number conservation. If this intimate connection is broken somehow, the phenomenology becomes very different.
Several mechanisms can violate KK number conservation, thereby allowing KK modes to decay into one another. 

A possibility is the inclusion of higher-dimensional KK-number-breaking operators in the bulk~\cite{Mohapatra:2003ah}.
In this case, new couplings become possible that connect modes in such a way that the decays are not forbidden by the phase space, and both the lifetime and the branching ratios will depend on the structure and the strength of these operators. 
This idea has been pursued in 5D in the dark dimension program by e.g. \cite{Montero:2022prj,Anchordoqui:2022svl,Gonzalo:2022jac,Obied:2023clp} which assumes that some breaking allows decays into a range of modes that almost fulfil KK-conservation and are phase-space allowed \footnote{Recently, Refs.~\cite{Lee:2026myr,Langhoff:2026wiz} raised some doubts regarding the width assumed by the aforementioned works. Concretely, they point out a misidentification of the relevant energy scale in the width estimate. Once corrected, this indicates that the strong enhancement of KK-graviton decays to lighter modes that is necessary to avoid decays to the SM is not realised. This has no implications for our work as KK-vectors do not couple to the SM directly and the decays via intermediate KK-gravitons are suppressed by higher powers of $\bar{M}_{Pl}$ as discussed in Sec.~\ref{sec:results}.}. 

Alternatively, one could include curvature in the extra dimensions, which breaks translation invariance in the extra-dimensional bulk and thus KK conservation. However, the way new channels open is non-trivial. In such a case, the mass relations do not necessarily open new kinematic channels. Rather, curvature loosens the KK selection rule, allowing new decay channels through a non-zero overlap function for modes which were \textit{already} kinematically open in the unperturbed model.
Explicit examples in 5D can be found both in the large-curvature regime (see e.g. Ref.~\cite{deGiorgi:2020qlg}) and in the small-curvature regime (see App.~\ref{app:small-curvature}) of the Randall-Sundrum model~\cite{Randall:1999ee,Randall:1999vf}.
In the two limits, the mass difference relevant for a decay $n_1\to n_2+n_3$ reads
\begin{equation}
    m_{n_1}-(m_{n_2}+m_{n_3}) \approx
    \begin{cases}
        (n_1-n_2-n_3)\frac{1}{R}\left(1-kR\dfrac{\pi}{2}\right)\,, & kR\ll1\,,\\[2mm]
        \pi k\, e^{-kR\pi}\left[(n_1-n_2-n_3)-\dfrac14\right]\,, & kR\gg1\,.
    \end{cases}
\end{equation}
In either regime, the sign of the left-hand side is determined by the integer combination $n_1-n_2-n_3$, as in flat space, yielding the same phase-space selection rule.
On the other hand, curvature opens new, previously forbidden, interactions among KK states. At $k=0$, KK number conservation forces $a_{n_1n_2n_3}=0$ if $n_1\neq n_2+n_3$. At $\mathcal{O}(kR)$, this overlap becomes non-zero. For instance, let us consider $n_1=3, n_2=n_3=1$: this satisfies $3>1+1$ and has $a_{3,1,1}=0$ at $k=0$, but turning on curvature gives (cfr.~App.~\ref{app:small-curvature})
\begin{equation}
    a_{3,1,1}=-\frac{32\sqrt{2}}{75\,\pi}\,kR+\mathcal{O}[(kR)^2]\approx -0.192\,(kR)\neq0\,.
\end{equation}
The analogue counterpart case in the large curvature limit was studied in detail, e.g., in Ref.~\cite{deGiorgi:2021xvm}, and yields the same qualitative conclusions.

\bigskip
We expect that the KK modes preferentially decay into the kinematically allowed channels that are closest to fulfilling KK-number conservation.
In practice, these are a small portion of the available states. In fact, the inequality of Eq.~\eqref{eq:mass-inequality} is saturated to $m_\si=m_\sj+m_\sk$ only when the angle between $\sj$ and $\sk$ is $|\theta_\sj-\theta_\sk|=0,\pi$, for $\si=|\sj\pm\sk|$, respectively. 
Therefore, we assume that a state $\si$ can decay in first approximation only into a pair of quasi-parallel or anti-parallel KK final states. In both cases, the vector $\vec{i}$ is approximately parallel to one of the decaying KK states; we can choose without loss of generality to be $\sj$, making the process kinematically possible whenever $\theta_\si\approx \theta_\sj$. Nevertheless, explicit computation reveals that the squared amplitudes of $V_i$ into arbitrary KK final states $X\in\{\phi,V,h\}$ scale as
\begin{align}
    &|\mathcal{M}(V_i\to X_j X_k)|^2\propto \sin(\theta_\si-\theta_\sj)^4\,,
\end{align}
which vanishes for $\theta_\si=\theta_\sj$. This implies that the closest available channels that the KK-violating parameters would not suppress are still closed due to the structure of the amplitude. 

Given the above result, it is clear that some tradeoff must be present.
To overcome this, we adopt the following strategy. We imagine we can open a new channel valid for $\si=\sj+\sk+\delta\sk$, where $\delta\sk$ is a small vector which deforms the sum rule. The perturbation $\delta\sk$ depends on its modulus $|\delta\sk|$ and an angle $\delta\theta$. 
In practice, we cannot know without the explicit model how $a_{\si\sj\sk}$ depends on such parameters; for example, some curvature might favour one direction over another. We remain agnostic and adopt a heuristic approach.
We define the bookkeeping index $\epsilon \ll 1$ which we use to take appropriate limits. We parametrise $\theta_j=\theta_i+\epsilon \Delta$, and assign an $\epsilon$ factor to the size of the perturbation $\epsilon|\delta\sk|$. Doing this, the phase space element depends on
\begin{equation}
    m_\si-(m_\sj+m_{\sk+\delta\sk})=\epsilon |\delta\sk|\cos(\theta_i-\delta\theta)+\mathcal{O}(\epsilon^2)\,.
\end{equation}
Therefore, the phase space is maximised when $\delta\theta \approx \theta_i$. We therefore also approximate in the following $\delta\theta=\theta_i+\epsilon \Delta_2$. In this way, all quantities that are required to be small do not have hierarchies among them.
Applying this prescription gives the amplitudes
\begin{align}
    &|\mathcal{M}(V_i\to V_j +\phi_k~\text{or}~h_k)|^2\propto \epsilon^2\frac{|\delta\sk|^2}{(m_\si-m_\sj)^2}\,, &|\mathcal{M}(V_i\to \text{ rest})|^2\propto \epsilon^3 \frac{\Delta^2|\delta\sk|}{(m_\si-m_\sj)}\,. 
\end{align}
This result highlights two things. First, assuming all small parameters are of the same order, a vector preferably decays into another vector and either a scalar or a graviton. Secondly, the amplitude is somewhat enhanced if $m_\sj \sim m_\si$. 

While it is impossible to draw firm conclusions without the full model, all in all, this suggests that the number of graviphotons is conserved during the decays. This will be our KK number-violation benchmark.


\subsection{Production Cross-Sections}
\label{sec:cross-section}
\begin{figure}
    \centering
    \includegraphics[width=\linewidth]{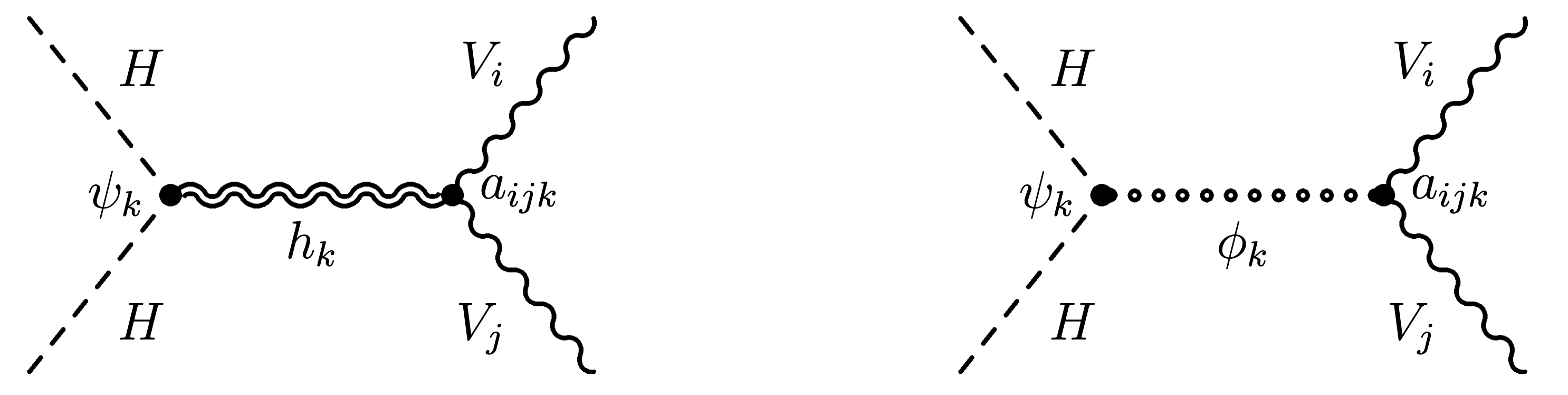}
    \caption{Representative set of diagrams contributing to the process $HH\to V_i V_j$.}
    \label{fig:diagrams}
\end{figure}

In this section, we consider the production of $V_iX_j$ from the SM bath.
We first study the scattering process of two scalars $HH\to V_i X_j$ where $X_j\in\{\phi_j,V_j,h_j\}$. Given the grid of non-redundant KK-modes and the selection rules for the $a_{\si\sj\sk}$ coefficients, there are only four Feynman diagrams contributing, which are a double copy with intermediate states differing only in the associated 2D KK-labels: two involving the exchange of gravitons and two with massive radions. The two types of diagrams for the $V_i V_j$ final state are shown in Fig.~\ref{fig:diagrams}.

In the high-energy limit where $s\gg m_H^2$, all SM content is essentially massless, and the contribution from the trace of the energy-momentum tensor $T=\eta^{\mu\nu}T_{\mu\nu}$ in Eq.~\eqref{eq:couplings} vanishes at tree-level due to the emerging conformal invariance. 
In other words, in such a limit, the contributions from the exchange of $\phi$-modes are negligible, and only diagrams with graviton exchanges are relevant. Nevertheless, during the computations, we included all diagrams and isolated the $\phi$-contributions to verify this explicitly to further assess the goodness of the results.

A crucial aspect of theories with massive higher-spin states is the behaviour of scattering amplitudes at high energies. The longitudinal polarisation modes of massive spin-2 gravitons and massive spin-1 vectors give rise to individual Feynman diagram contributions that grow as powers of $s$, threatening perturbative unitarity already at relatively low scales. Restoring unitarity requires the inclusion of the full spectrum of KK states generated upon integrating out the extra dimensions, and the exact cancellation mechanism depends on the geometry of the extra dimensions~\cite{Schwartz:2003vj,Bonifacio:2019ioc,Bonifacio:2019mgk}.
This has been verified explicitly in extra-dimensional models~\cite{Chivukula:2020hvi,Chivukula:2023qrt,Chivukula:2024vgg}, where a delicate series of cancellations takes place among the full set of Feynman diagrams requiring the participation of the complete KK particle content. Analogous results have been established in the presence of matter fields~\cite{deGiorgi:2020qlg,Chivukula:2023sua,deGiorgi:2023mdy}, where the interplay between gravitational and matter interactions proves equally non-trivial.

The present framework contains a richer field content, with KK scalars and vectors alongside KK gravitons. Nevertheless, the same mechanism is expected to operate here. Indeed, we find that individual diagrams involving longitudinal gravitons and vectors yield contributions that grow as $s^2$, consistent with this picture. For the non-trivial cancellations to take place and cancel such terms, it is required that the full set of diagrams involving gravitons, graviphotons and scalars, as well as all the non-redundant allowed KK states, are carefully taken into account. Since this work is mostly dedicated to phenomenology, we report here the main findings useful for DM, and leave further details in  App.~\ref{app:high-energy} for the interested reader. 

\bigskip
In the high-energy limit, the matrix element squared can be obtained straightforwardly by means of Eq.~\eqref{eq:helicity-squared}. The amplitudes for the three cases read
\begin{align}
    &|\mathcal{M}(HH\to V_i \phi_j)|^2\approx (\psi_k a_{ijk}\sin(2\Delta))^2\frac{\sin^4\theta}{6\bar{M}_{Pl}^4}s^2\,,\\
    &|\mathcal{M}(HH\to V_i V_j)|^2\approx (\psi_k a_{ijk}\cos(2\Delta))^2\frac{\sin^4\theta}{4\bar{M}_{Pl}^4}s^2\,,\\
    &|\mathcal{M}(HH\to V_i h_j)|^2\approx (\psi_k a_{ijk}\sin(2\Delta))^2\frac{\sin^4\theta}{12\bar{M}_{Pl}^4}s^2\,,
\end{align}
where $\Delta=\theta_i-\theta_j$.
As can be seen, the squared amplitudes grow with $s^2$, the minimum order required by the two gravitational vertices in these processes.
The total probability of producing a vector $V_i$ with an associated $X_j\in\{\phi_j,V_j,h_j\}$ particle is then given in the high-energy limit by
\begin{align}
\label{eq:sigma:ij}
|\mathcal{M}(V_i)|^2&=\sum\limits_{X_j}|\mathcal{M}(HH\to V_i X_j)|^2\\
&\nonumber\approx (\psi_k a_{ijk})^2\frac{\sin^4\theta}{\bar{M}_{Pl}^4}s^2\left[\frac{\cos^2(2\Delta)}{4}+\sin^2(2\Delta)\left(\frac{1}{6}+\frac{1}{12}\right)\right]
=(\psi_k a_{ijk})^2\frac{\sin^4\theta}{4\bar{M}_{Pl}^4}s^2\,.
\end{align}
Remarkably, all factors, depending on the type of KK states produced, disappeared. This is valid only for the leading order in the large-$s$ expansion.  The dependence on the moduli of $i,j$ reappears at $\mathcal{O}(s^1)$ while their angular dependence $\Delta$ reappears at $\mathcal{O}(s^0)$.

\bigskip
The amplitudes presented above include only scalar initial states. For phenomenology, all SM states are relevant.
For comparison, we present the amplitudes involving different SM states into a pair of $V_i V_i$. The amplitudes squared and summed over initial and final spins for different processes at leading order in the large-$s$ expansion are:
\begin{align}
    \sum\limits_\text{spins}|\mathcal{M}(HH\to V_i V_i)|^2 &\approx  \frac{\psi_0^2    a_{0ii}^2 s^2  \sin ^4(\theta )}{4 \bar{M}_{Pl}^4}\,,\\
    \sum\limits_\text{spins}|\mathcal{M}(ZZ\to V_i V_i)|^2 &\approx \frac{a_{0ii}^2 \psi_0^2  s^2\sin ^4(\theta )}{2 \bar{M}_{Pl}^4} \,,\\
    \sum\limits_\text{spins}|\mathcal{M}(\gamma\gamma\to V_i V_i)|^2 &\approx \frac{a_{0ii}^2 \psi_0^2  s^2\sin ^4(\theta )}{2 \bar{M}_{Pl}^4}\,,\\
    \sum\limits_\text{spins}|\mathcal{M}(\bar{\psi}\psi\to V_i V_i)|^2 &\approx \frac{2 a_{0ii}^2 \psi_0^2 s^2 \sin ^2(\theta ) \cos ^2(\theta )}{\bar{M}_{Pl}^4}\,,
\end{align}
where $\psi$ is any SM fermionic singlet under QCD. Again, subtle cancellations between all relevant Feynman diagrams take place, but the qualitative picture remains the same and all the squared matrix elements grow with $s^2$ in the high-energy limit (see App.~\ref{app:high-energy}).

The results for other initial states, such as quarks, gluons, or $W$-bosons, can be obtained by including the appropriate spin-averaging factors, multiplying by the respective colour and multiplicity factors and adjusting the factor $1/2$ for indistinguishable particles.
In the same limit, the cross-section per channel can be written as
\begin{equation}
    \sigma(XX\to V_i V_i)\approx  \frac{c_X}{240\pi}\,(a_{0ii}  \psi_0)^2\left(\frac{s}{\bar{M}_{Pl}^4}\right)\,,
\end{equation}
where the initial state-dependent coefficients read
\begin{align}
    \label{eq:coefficients}&c_H=1\,, &&c_Z=\frac{2}{9}\,, &&c_\gamma=\frac{1}{2}\,, &&c_\psi=\frac{1}{2}\,.
\end{align}

Before moving on, it is interesting to note that the KK production processes are better behaved than  KK-mediated SM processes. 
As the KK-selection rules do not restrict the number of excitations appearing in the internal propagator 
Tree-level processes which involve the exchange of KK modes can become divergent. 
In $d=1$  the problem is merely formal as the sum of propagators converges, and thus observables are automatically insensitive to the cut-off of the theory. However, in $d\geq 2$ this issue persists, as can be seen straightforwardly when considering two-to-two scattering of matter states via s-channel graviton mediation. In $d\geq 2$ dimensions, the amplitude is proportional to
\begin{equation}
    i\mathcal{M}\propto i\mathcal{M}_0 \times \sum\limits_{\vec{n}_d}\frac{1}{s-m_{\vec{n}_d}^2}\sim \sum\limits_{\vec{n}_d}\frac{1}{|\vec{n}_d|^2}\sim \int\limits_1^N dn~\frac{n^{d-1}}{n^2}\sim \begin{cases}
        \log N & d=2\,,\\
        N^{d-2} & d\geq 3\,.
    \end{cases}
\end{equation}
Thus, this matrix element diverges logarithmically for $d=2$ and polynomially for $ d\geq 3$ extra dimensions, while the matrix elements for KK-graviton production or scattering remain finite.

\section{Production in the early Universe}
\label{sec:early-universe}

As all graviphoton couplings are Planck-suppressed in this model, interaction rates are naturally very low, and no dark matter components equilibrate with the SM plasma. Thus, DM is produced by a freeze-in mechanism~\cite{Hall:2009bx}. In our computation, we assume KK states to be much heavier than all the SM ones. This will turn out to be justified based on our results. For simplicity, we further adopt an instantaneous reheating prescription, and take the SM to be in equilibrium at a reheating temperature $T_r$, see e.g. Refs.~\cite{Mambrini:2021zpp,Henrich:2024rux} for a discussion of a gravitational freeze-in with a more careful modelling of reheating.  The KK sector is not yet populated at this stage, and accumulates via two-to-two scattering of SM states.
Within this context, the most interesting physics case lies when the model satisfies the hierarchy of scales $Nm_1 \gg T_r \gg m_1$. In this way, the EFT nature of the theory is not relevant for our predictions, and the observables become insensitive to the artificial cut-off of the KK towers, $N$.

The production of the vector can be described by a Boltzmann equation. Neglecting the back-reaction term and using classical statistics for simplicity, one finds 
\begin{align}
    \label{eq:boltzmann}\frac{d n_V}{ d t}+3 H n_V = \sum\limits_{a \in \text{SM}} 2 \langle\sigma v\rangle_{2V}\, n_{\text{a,eq}}^2  +  \langle\sigma v\rangle_{1V}\, n_{\text{a,eq}}^2 \,,
\end{align}
where $H$ is the Hubble rate, while $\langle \sigma v \rangle_{2V (1V)}$ is the thermally averaged pair- (single-)production cross section of the vector model and $n_{\text{a,eq}}$ is the equilibrium density of the SM particle species $a$.
Neglecting quantum statistics effects, the terms on the left-hand side can be computed as \cite{Gondolo:1990dk}
\begin{align}
   &\label{eq:sigmavn2} \langle\sigma v\rangle_{ij}\, n_{\text{a,eq}}^2=  \frac{g_{a}^2 T}{32 \pi^4} \int^{\infty}_{(m_i+m_j)^2} ds~ \sigma_x(s) s^{3/2} K_1\left( \frac{\sqrt{s}}{T}\right)\,,
\end{align}
where $g_{a}$ counts the internal degrees of freedom of the respective SM particles, and $m_{i,j}$ are the KK masses of the final states, while $K_1$ denotes the first modified Bessel function of the second kind. Here we anticipate that all incoming particles are relativistic in the regime of interest here.

The structure of $\sigma_x$ is highly non-trivial, and we will need some simplifications to allow for an efficient computation of the relic density. Let's first consider the high-energy limit. 
Here the production cross-section is independent of the SM initial states up to a numerical prefactor, so that we can approximate 
\begin{align}
    \label{eq:rescaling} 
    &\sum\limits_{x\in\text{SM}}g_x^2\sigma_x(s)\approx  c_\text{SM}~\sigma_{HH}(s)\,,
\end{align}
where $c_\text{SM}= \sum g_x^2c_x=70$.
We exploit this by using the cross section for the production from scalar particles as a template and generalise to the full SM by applying the rescaling of Eq.~\eqref{eq:rescaling} in all kinematic regimes. 
As we will see in the next section, the final result will be highly insensitive to $\mathcal{O}(1)$ factors, which justifies the use of the above approximation.

\bigskip
Finally, in the following we will work in a radiation-dominated Universe such that the Hubble parameter, the entropy density and the equilibrium number density of a classical particle are given by 
\begin{align}
    &H=\sqrt{\frac{\pi^2 g_\star}{90}}\frac{T^2}{\bar{M}_{Pl}}\equiv c_H\frac{T^2}{\bar{M}_{Pl}}\,, &&s=\frac{2\pi^2g_s}{45} T^3\equiv c_s T^3\,, &&n_{\text{Bolz,eq}}=\frac{T^3}{\pi^2}\equiv c_n T^3\,,
\end{align}
where, for future convenience, we factorised out dimensionful quantities and defined the dimensionless numerical coefficients by $c_{H,s,n}$. 

\subsection{Yield Estimate}
\label{sec:yield}

The relic abundance of a single 
dark matter species $X$ is conveniently  obtained by tracking the evolution of its 
yield $Y = n/s$, where $n$ is the number density and $s$ is the 
entropy density. The yield satisfies the Boltzmann equation~\eqref{eq:boltzmann}, which 
in the freeze-in regime can be written at temperatures above electroweak-weak symmetry breaking as
\begin{equation}
    -\frac{dY}{dT} = \sum\limits_{a \in \text{SM}} \frac{\langle\sigma v\rangle\, n_{\text{a,eq}}^2}{T\, H\, s}\,,
\end{equation}
where $\langle \sigma v \rangle $ is to be understood as the properly weighted sum of the single and double production cross sections of the mode under consideration. Its relic abundance today is then given by
\begin{equation}
    \Omega_0 h^2 = -\frac{s_0\, m_X}{\rho_c/h^2}\, \int\limits_0^{T_r}dT~\frac{dY}{dT} \,,
\end{equation}
where $s_0$ is the present entropy density and $\rho_c$ is the 
critical density. 

This generalises straightforwardly to the multi-field case, where 
one must sum over the full tower of KK vector states $V_{\vec{i}}$. 
The total relic abundance in a stable tower is obtained by solving 
\begin{align}
\label{eq:sum_Yi}
   -\sum_{\vec{i}} m_{\vec{i}}\, \frac{dY_{\vec{i}}}{dT}
   \approx \sum\limits_{a \in \text{SM}} \sum_{\vec{i}\vec{j}}
    \frac{m_{\vec{i}}\, n_{\text{a,eq}}^2}{T\,H\,s}\,\langle\sigma v\rangle_{a,\vec{i}\vec{j}}
    \approx 
    \left(\frac{\bar{M}_P c_n^2 c_{\text{SM}}}{c_s\, c_H}\right)
    \sum_{\vec{i}\vec{j}} m_{\vec{i}}\,\langle\sigma v\rangle_{\vec{i}\vec{j}}\,,
\end{align}
where in the last step we have used the standard radiation-dominated 
relations to simplify the prefactor. 
In the $N\gg 1$ limit, the sums over KK-states can be traded for integrals
\begin{equation}
     \sum_{\vec{i}\vec{j}} (\dots) \longrightarrow \pi\int_{1}^N di \, i~\int_{1}^N dj j\int\limits_{-\pi/2}^{\pi/2} d\Delta~ (\dots)\,.
\end{equation}
where we used the fact that the observables depend only on the angular difference $\Delta=\theta_i-\theta_j$.
Applied to the case of interest, the total abundance can then be computed via
\begin{equation}
    \Omega_{\text{DM},0}h^2\approx  \frac{s_0 \bar{M}_P}{\rho_c/h^2}\left(\frac{\pi c_n^2 c_{\text{SM}}}{c_s c_H} \right) m_1 \times S(T_r,N)\,,
\end{equation}
where
\begin{equation}
    S(T_r,N)\equiv \int\limits_{0}^{T_r}dT\int_{1}^N di~i^2~\int_{1}^N dj j\int\limits_{-\pi/2}^{\pi/2} d\Delta~\langle\sigma v\rangle_{ij}(T)\,,
    \label{eq:S-def}
\end{equation}
and $N$ denotes the number of the highest mode we want to consider. As discussed below, the concrete choice of $N$ will not be relevant in the end provided $m_N$ is sufficiently higher than the reheating temperature $T_r$. 

In models where the tower is unstable and decays, the contribution to the relic density depends on the structure of the decays. For the cosmologically fast vector number conserving decays we identified as our benchmark scenario in the previous section, one just needs to sum up the $Y_i$ and apply a universal weight of $m_1$ instead of $m_i$ in Eq.~\eqref{eq:sum_Yi}, which amounts to dropping one power of $i$ in Eq.~\eqref{eq:S-def}.

\bigskip
We compute the contribution to the total abundance in three different regimes: (i) fully relativistic, $m_{\vec{i}}\ll T$, (ii) fully non-relativistic, $m_{\vec{i}} \gg T$, and (iii) intermediate, $m_{\vec{i}} \sim T$. The non-relativistic regime (ii) is exponentially suppressed relative to the relativistic one (i) and therefore plays no significant role in our results. The intermediate regime (iii), on the other hand, gives the dominant contribution. This can be understood as follows: the number of accessible KK modes grows as $N^2$, so the bulk of the producible spectrum is concentrated near the highest accessible energy, $m_{\vec{i}} \sim T$, precisely where the intermediate regime applies.

\subsubsection{Relativistic Modes}
\label{sec:relativistic}

In this regime, the cross section is particularly simple and can be parametrised as $\sigma(s) = C\, s / \bar{M}_P^4$, such that for $T \gg m_\si$:
\begin{align}
    &\langle\sigma v\rangle_{\vec{i} \vec{j}}\,  \approx c_\text{SM}~24C_{\vec{i} \vec{j}}\left(\frac{T^2}{\bar{M}_P^4}\right)\,.
\end{align}
As expected, production is dominated by the highest energies, and corrections relevant for $T \approx m_{V_{\vec{i}}}$ do not matter as long as $T_r \gg m_{V_{\vec{i}}}$. An analogous result holds for single vector production.
Recalling the matrix elements from Eq.~\eqref{eq:sigma:ij}, we identify
\begin{equation}
    C_{\vec{i}\vec{i}} = \frac{(a_{0\vec{i}\vec{i}}\,\psi_0)^2}{240\pi} = \frac{1}{240\pi}\,,\qquad
    C_{\vec{i}\vec{j}} = \frac{(\psi_\sk\, a_{\vec{i}\vec{j}\sk})^2}{120\pi} = \frac{1}{120\pi}\,.
\end{equation}
The coefficient $C_{\vec{i}\vec{j}}$ (with $\vec{i} \neq \vec{j}$) contains all possible final states, including vectors. To obtain the contribution to population $\vec{i}$, one sums over $\vec{j}$, replacing the sum by an integral in the $N \gg 1$ limit:
\begin{equation}
    C_{\vec{i}} \equiv \sum_{\vec{j}} C_{\vec{i}\vec{j}}
    \approx \frac{1}{120\pi} \times \frac{\pi}{2}N_\text{rel}^2
    = \frac{N_\text{rel}^2}{240}\,,
\end{equation}
where 
$N_\text{rel}$ is the number of the highest KK modes for which one wants to use the relativistic approximation. It has to satisfy $N_\text{rel}/R \lesssim T_r$. If $N_\text{rel} \gg 1$, the single-vector production channel dominates. The Boltzmann equation for the yield can then be written as
\begin{equation}
    -\frac{dY_{\vec{i}}}{dT}
    \approx \sum_a \frac{\langle\sigma v\rangle_{V_{\vec{i}}}\, n_{\text{a,eq}}^2}{T\,H\,s}
    \approx \frac{c_\text{SM}}{10}\frac{c_n^2}{c_s\, c_H}
    \frac{N_\text{rel}^2\, T^2}{\bar{M}_P^3}\,.
\end{equation}
Integrating from $0$ to $T_r$ one finds the yield of species $V_\si$ to be
\begin{equation}
    Y_{\vec{i}} \approx \frac{c_\text{SM}}{30}\frac{c_n^2}{c_s\, c_H}
    \frac{N_\text{rel}^2\, T_r^3}{\bar{M}_P^3} \equiv Y\,,
\end{equation}
which is universal across all relativistic species. 

The total DM abundance is obtained by summing these contributions with the appropriate weight as determined by the stability assumptions of these modes.
For example, if the tower is stable, the total relic abundance receives contributions from relativistic modes that are weighted by their mass. With 
\begin{equation}
    \sum_{\vec{i}} m_{\vec{i}}
    \approx \pi m_1\times\int_1^{N_\text{rel}} dn\, n^2
    \approx \frac{\pi\, N_\text{rel}^3}{3}m_1\,,
\end{equation}
we therefore find
\begin{equation}
    \label{eq:omega-relativistic}\Omega_{\text{DM},0}h^2
    = \frac{s_0}{\rho_c/h^2}\frac{c_\text{SM}}{90}
    \frac{\pi c_n^2}{c_s\, c_H}\frac{N_\text{rel}^5}{\bar{M}_P^3}m_1T_r^3 = \frac{s_0}{\rho_c/h^2}\frac{c_\text{SM}}{90}
    \frac{\pi c_n^2}{c_s\, c_H}\,\frac{\epsilon^5\, T_r^{8}}{m_1^4\,\bar{M}_P^3} \,,
\end{equation}
where in the last step we identified $N_\text{rel} = \epsilon T_r/m_1$ with $\epsilon \lesssim 1$.
As expected, the result is strongly UV-dominated with a $T_r^8$ dependence.

\subsubsection{Non-Relativistic Modes}
\label{sec:non-relativistic}

Let us now turn to the non-relativistic contribution. If $m_N\gg T_r$, modes with $m_\si\gg T_r$ will be in the non-relativistic regime for the entire thermal history, i.e. $\sqrt{s_\text{min}}=(m_i+m_j)\gg T$. In this limit, one can make use of a different formalism, as the vector production is thermally suppressed and only the cross section on-threshold matters. 
The relevant matrix elements squared for the process $HH\to V_i X_j$ on threshold  read
\begin{align}
    &|M(HH\to V_i \phi_j)(s_\text{min})|^2=0\,, \\
    &|M(HH\to V_i V_j)(s_\text{min})|^2=2(a_{ijk}\psi_k)^2\frac{s_\text{min}^2}{\bar{M}_P^4}\,, \\
    &|M(HH\to V_i h_j)(s_\text{min})|^2=0\,.
\end{align}
Therefore, near the threshold, only vector pairs are efficiently produced.

In the non-relativistic limit, the cross-section can be approximated as
\begin{equation}
\label{eq:non-rel-sigma}
   \sigma(s_\text{min} w^2)= (a_{ijk}\psi_k)^2\frac{(m_im_j)^{1/2}(m_i+m_j)}{4\pi \bar{M}_P^4}(w^2-1)^{1/2}+\mathcal{O}((w^2-1)^{3/2})\,,
\end{equation}
where we defined $w^2\equiv s/s_\text{min}$ and the $(w^2-1)^{1/2}$ factor accounts for the phase space closure.
The integral for the thermal average of Eq.~\eqref{eq:sigmavn2} can be solved in this case. The leading contribution is given
\begin{align}
       \langle\sigma v\rangle_{V_i}\, n_{\text{a,eq}}^2\approx &\sum_j \frac{ T^3}{128 \pi^{4} \bar{M}_P^4}(m_im_j)^{1/2}(m_i+m_j)^4 e^{\left(-\frac{m_i+m_j}{T}\right)}\,.
    \end{align} 
    As expected, this contribution exhibits the exponential Boltzmann suppression, and thus modes with $m_i\gg T$ contribute negligibly to the relic density despite the large number of heavy modes.

\subsubsection{Intermediate regime}
\label{sec:intermediates}

Finally, we turn to the intermediate regime. As the cut-off of the theory has to be above the reheating temperature in order to allow for a consistent computation, some modes will not live exclusively in the relativistic or non-relativistic regimes, but rather in the intermediate regime. While non-relativistic modes lie in the Boltzmann-suppressed tail, their quadratic growth in number competes with the onset of the $e^{-(i+j)m_1/T}$ suppression. Therefore, the expectation that modes with $m_i\gtrsim T$ do not contribute significantly is too naive, while the expectation remains valid whenever $m_i\gg T$. 
 
Let us denote by $N$ the modulus of the largest KK vector included in the computation. Ideally, with unlimited computational power, one could take $N$ to be as large as the cut-off. However, this is not necessary, as the Boltzmann suppression ensures that such heavy modes do not contribute for $T_r$  below the cut-off. Thus we then keep $N$ as a free parameter and investigate how many modes are needed to obtain a meaningful result.

The key quantity that we need to compute in order to estimate the total DM abundance is $S(T_r,N)$, as defined in Eq.~\eqref{eq:S-def}. From dimensional analysis,  its scaling is expected to be $T_r^8/m_1^5\bar{M}_{\rm Pl}^4$, in analogy to the result found for relativistic modes in Eq.~\eqref{eq:omega-relativistic}.  However, we have not been able to find a simple closed expression for its numerical prefactor, as neither the relativistic nor non-relativistic expansions provide an acceptable fit to the full cross-section. Therefore, an approximation has to be used.

\begin{figure}
    \centering
    \subfigure[]{
        \includegraphics[width=0.495\linewidth]{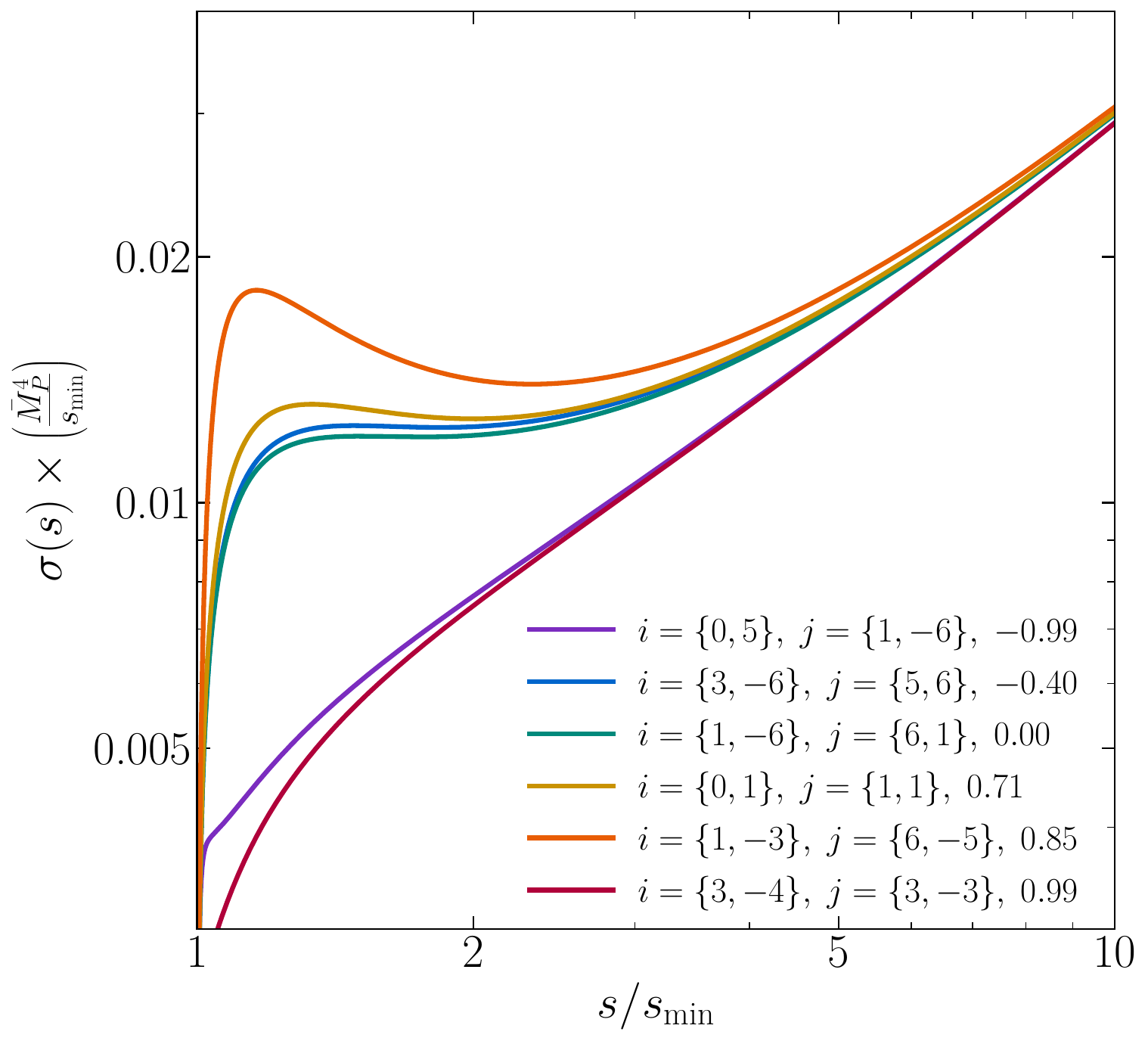}
        \label{fig:sigma_cross}
    }
    \subfigure[]{
        \includegraphics[width=0.455\linewidth]{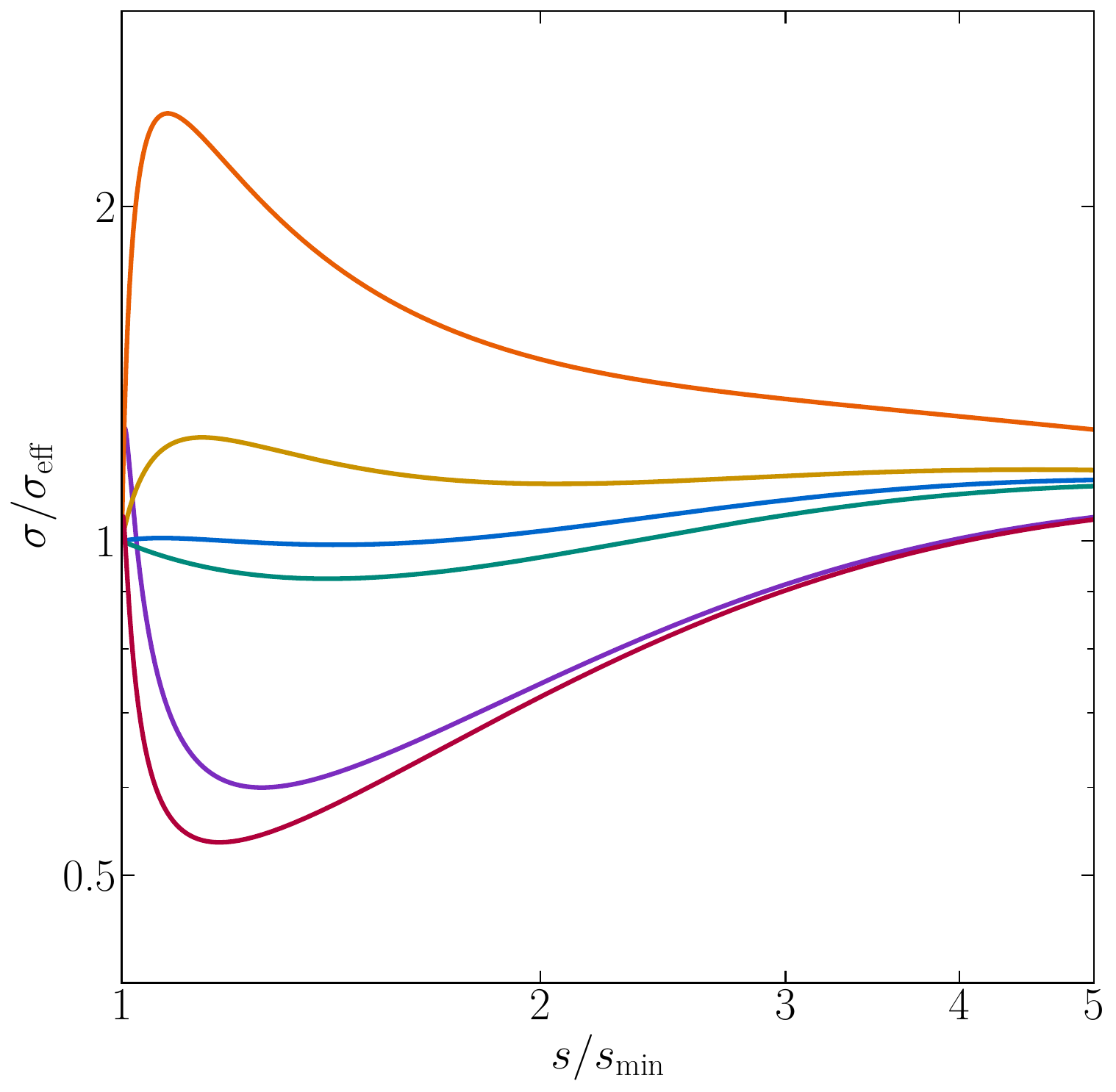}
        \label{fig:sigma_ratio}
    }
    \caption{The legend is shared between the two plots. Each colour corresponds to a choice among a representative set of $\si,\sj$ vectors; the corresponding cosine of the angle difference $\cos\Delta$ is shown also shown.  \textbf{Left.} Energy dependence of the exact cross section $\sigma(s)$. \textbf{Right.} Ratio of the exact cross-section to the effective approximation
    $\sigma_{\rm eff}$ of Eq.~\eqref{eq:sigma-effective} as a function of $s/s_{\rm min}$.}
    \label{fig:sigma_comparison}
\end{figure}

\bigskip
Our strategy is to simplify the expression for the cross-section by finding an appropriate surrogate, and to bracket the final result with suitable upper- and lower-estimates. This procedure is then verified by a numerical study in the low $T_r$ limit where a modest number of modes is sufficient, see below for more details.
In order to build a suitable effective cross-section, we note that the exact amplitude receives contributions from two $s$-channel diagrams of Fig.~\ref{fig:diagrams} with $\vec{k}=|\vec{i}+\vec{j}|$ and $\vec{k}=|\vec{i}-\vec{j}|$ which are proportional to $1/s_+$ and $1/s_-$, where
\begin{equation}
    s_\pm \equiv s - |\vec{m}_i\pm\vec{m}_j|^2 = 
    s - \left(m_i^2 + m_j^2 \pm 2m_im_j\cos\Delta\right)\,,
\end{equation}
with $m_i=|\vec{m}_i|$ and $\Delta$ the angle between $\vec{m}_i$ and $\vec{m}_j$. These diagrams develop an enhancement just above the kinematic threshold $s_{\rm min}=(m_i+m_j)^2$ as $s_\pm\to 0$. Despite the divergence in the propagator, the amplitude self-regularises and keeps the result finite, leaving a finite enhancement near the kinematical threshold. 
Such an effect can be seen in the left panel of Fig.~\ref{fig:sigma_comparison}.
We incorporate this effect by retaining the full $s_\pm$ dependence in the bracket, with $1\pm \cos\Delta$ coefficients fixed by requiring the correct behaviour at both asymptotic limits. The effective cross-section we consider is then
\begin{align}
    \label{eq:sigma-effective}\sigma_{\rm eff}(s) \equiv\; &
    \frac{1}{120\pi\bar{M}_{\rm Pl}^4}
    \left(s-(m_i-m_j)^2\right)^{1/2}
    \left(s-(m_i+m_j)^2\right)^{1/2}\times
    \\
    &\nonumber\qquad\qquad\qquad\qquad\times\left[1 + 14\frac{(m_i+m_j)^2}{s}
    \times(m_im_j)\left(\frac{1+\cos\Delta}{s_-}
    +\frac{1-\cos\Delta}{s_+}\right)\right]\,.
\end{align}
The numerical coefficients $1/120$ and $14$ are fixed by requiring $\sigma_{\rm eff}$ to reproduce the correct behaviour at the kinematic threshold of
Eq.~\eqref{eq:non-rel-sigma}, where $s\to s_{\rm min}$, and in the high-energy limit of Eq.~\eqref{eq:sigma:ij}, where $s\to\infty$ and
$\sigma_{\rm eff}\approx s/(120\pi\bar{M}_{\rm Pl}^4)$. 

To assess the quality of this approximation, we compare it to the exact numerical result for a representative set of mode pairs in Fig.~\ref{fig:sigma_comparison}. The ratio $\sigma/\sigma_{\rm eff}$ is shown as a function of $w^2\equiv s/s_{\rm min}$ for several representative choices of $m_{\si,\sj}$. As can be seen, all curves approach unity for $w^2\gg 1$ and $w^2\to 1$, as expected from the construction requirements of $\sigma_\text{eff}$. Near threshold, deviations become more pronounced. We checked for a large set of vectors that this could be as large as a factor of five for nearly parallel vectors ($\cos\Delta\approx 1$), where the collinear enhancement of the exact cross section is most pronounced. However, for $\cos\Delta=0$ the agreement is excellent across the full kinematic range.  This case corresponds to orthogonal vectors and is the most probable value for the cosine for randomly oriented modes. Since the mode sum is dominated by pairs with typical relative angles $\Delta\sim\pi/2$, we are confident that $\sigma_{\rm eff}$ provides a reliable approximation for the computation of $S(T_r,N)$.

\bigskip
Even though the expression for $\sigma_\text{eff}$ is already much simpler than the exact one, it is still not possible to obtain an analytical result. To derive upper and lower bounds, we replace the $\Delta$-dependent part of $\sigma_{\rm eff}$ with its maximum and minimum values over the integration range, which happen at $\Delta=\pi/2$ and $\Delta=0,\pi$, respectively. The procedure is described in App.~\ref{app:intermediate-yield} and yields the narrow range
\begin{equation}
    \frac{29\,T_r^8}{m_1^5\,\bar{M}_{\rm Pl}^4}
    \;\leq\; S(T_r,N) \;\leq\;
    \frac{42\,T_r^8}{m_1^5\,\bar{M}_{\rm Pl}^4}\,.
\end{equation}
Since most of the modes statistically lie near the maximum region where $\cos\Delta=0$, one expects the coefficient to be closer to $42$. A full numerical evaluation of the five-dimensional integral confirms that the exact result falls well within these bounds and expectations and indicates
\begin{equation}\label{eq:approx-S}
    S(T_r,N) \approx 37\frac{\,T_r^8}{m_1^5\,\bar{M}_{\rm Pl}^4}\,.
\end{equation}

\begin{figure}
    \centering
    \includegraphics[width=0.8\linewidth]{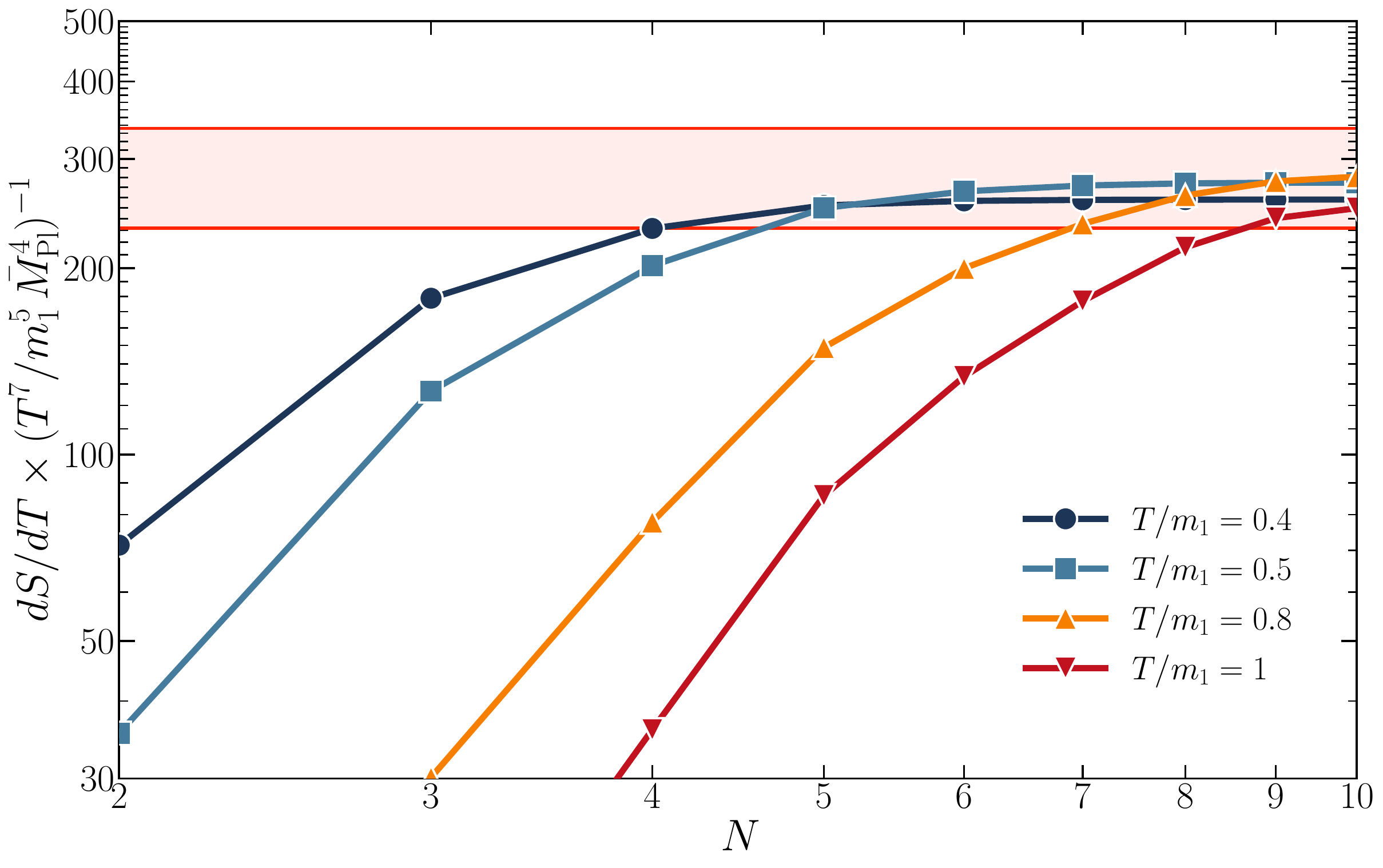}
    \caption{Comparison between the exact numerical results for different values of $T$ as a function of $N$ and the estimate stemming from the effective cross-section of Eq.~\eqref{eq:approx-S} (in between red-dashed lines).
    }
    \label{fig:S_comparison}
\end{figure}
The comparison between our analytical estimation of Eq.~\eqref{eq:approx-S} and the exact numerical result stemming from the full set of amplitudes is shown in Fig.~\ref{fig:S_comparison}. The plot shows the value of $dS/dT$ normalised to its expected scaling $dS/dT\propto T^7/(m_1^5 \bar{M}_\text{Pl}^4)$ for different temperatures as a function of the heaviest KK mode included in the sum, defined by the KK number $N$. The value of $S(T_r,N)$ quoted in Eq.~\eqref{eq:approx-S} can then be obtained by integrating over the temperature. As can be seen, as the value of $N$ increases, the curves do approach a plateau value which falls within the estimated range of Eq.~\eqref{eq:approx-S}. We can also read off that approximately the curves keep growing until the mode $N\approx 10~T/m_1$ is included within the sum. These correspond to modes which are already almost fully non-relativistic. Even heavier modes are too Boltzmann suppressed to contribute meaningfully to the total yield. All in all, this confirms that the approximation employed for the cross-section via Eq.~\eqref{eq:sigma-effective} is justified.

The estimation of total abundance is then
\begin{equation}
    \label{eq:omega-intermediate}
    \Omega_{\text{DM},0}h^2\approx  \frac{s_0 }{\rho_c/h^2}\left(\frac{37\pi c_n^2 c_\text{SM}}{c_s c_H} \right) \times \frac{\,T_r^8}{m_1^4\,\bar{M}_{\rm Pl}^3}\,.
\end{equation}
Compared to what was previously obtained from the relativistic modes in Eq.~\eqref{eq:omega-relativistic}, the intermediate contribution is three orders of magnitude larger. Thus, both the relativistic and non-relativistic modes can be safely neglected in the determination of the total dark matter relic density. 

The computation above was performed employing the definition of $S(T_r,N)$ from Eq.~\eqref{eq:S-def}, which assumes all graviphotons to be stable. For our other benchmark, the decaying tower with vector number conservation, the total yield is the relevant quantity. A similar computation leads to
\begin{equation}
    Y_{\text{DM},0}\approx  \left(\frac{\pi c_n^2 c_\text{SM}}{c_s c_H} \right)~\frac{T_r^7}{m_1^4 \bar{M}_P^3}\times \left[8,11\right]\,.
\end{equation}
Performing the sum over $N$ numerically with a similar strategy to the one mentioned above points towards a coefficient of  $\approx 10$, which we adopt for such a scenario.
Compared to the result for the stable tower of Eq.~\eqref{eq:omega-intermediate}, the expression grows with a lower power of $(T_r/m_1)$. This originates from the missing $m_\si$ factor when summing over the KK states.

\subsection{Constraints and Results}
\label{sec:results}

Equipped with the estimation of the yields from the previous section, we can now turn to the available parameter space. Before proceeding with the results, we need to take some constraints into account: these include the validity of the EFT, bounds from BBN, and limits on the DM lifetime from indirect detection.

\subsubsection{Constraints}

\paragraph{Validity of the EFT}
It is clear that our computation can only be expected to hold as long as the reheating temperature is smaller than the $D$-dimensional Planck mass $M_D$. Using the relation between $M_D$, $M_{Pl}$ and the mass of the lightest KK-mode this leads to
\begin{align}
    T_r \ll M_D= \sqrt{\frac{m_1\,\bar{M}_{Pl}}{2 \pi}}\,.
\end{align}

\paragraph{BBN.}
Naturally, the spin-2 modes of the KK tower can be produced by similar processes. As they couple directly to the stress-energy tensor, their decay width is drastically larger than that of the vector modes. Nevertheless, these decays are still Planck-suppressed and, therefore, the gravitons can be quite long-lived compared to the time scales of the early Universe. Once the lifetime of the lightest mode exceeds about 10 seconds, part of the KK population will decay during Big Bang Nucleosynthesis. This injects additional energy into the SM plasma, which can affect the abundance of the primordial elements.  

The width of the lightest KK graviton to SM final states is given by~\cite{deGiorgi:2022yha}
\begin{align}
    \Gamma_{1}=\frac{73}{240\pi} \frac{m_1^3}{\bar{M}_{Pl}^2}
\end{align}
where we have neglected the masses of SM particles. A conservative lower limit on $m_1$ can be found by requiring that this population has decayed before BBN becomes sensitive to additional energy injection. This corresponds to a lifetime of less than $10$ seconds \cite{Kawasaki:2017bqm} or $m_1\gtrsim 16$ TeV provided that the reheating temperature is high enough to populate these states.
If the mass is lower than this value, there will always be KK modes with a lifetime that matches the age of the Universe during BBN. However, this estimate is overly conservative. Since all the interactions are Planck-suppressed, the yield of the gravitons and, consequently, the amount of energy released in their decay is quite low. This was taken into account in Ref.~\cite{Gross:2024zdh}, which finds a significantly lower limit of just $\sim 300$ GeV based on a recast of Ref.~\cite{Kawasaki:2017bqm} for towers of KK-gravitons.

\paragraph{Vector lifetime and indirect detection.}
The vectors cannot decay to pairs of SM particles or the zero mode of the gravitational fields. However, this does not guarantee their absolute stability, and they can still decay into more complex final states, albeit at a higher power in $\bar{M}_{Pl}$.
Here we need to differentiate between two possibilities: 

\textit{1) KK-conservation holds:} In this case, the heavy modes in the tower are stable against inter-tower decays since the final states that are compatible with the KK-number selection in the couplings have a vanishing phase space as discussed in Sec.~\ref{sec:decay}. However, this does not ensure their absolute stability, and three-body decays into a lighter KK-state and a pair of SM particles from taking one KK-graviton off-shell remain possible at the price of a higher suppression in $\bar{M}_{Pl}$, as can be seen in Fig.~\ref{fig:decay-vectors}. The matrix element for this three-body process will parametrically be given by $|\mathcal{M}|^2\propto m_i^4/\bar{M}_{Pl}^4$. By combining this with the three-body phase space for a pair of massless particles and a massive particle, see e.g. \cite{Fuss:2022zyt}, this leads for a representative choice of $m_f=m_i/2 $ to a width of
\begin{align}
    \Gamma_{i,3-\text{body}}\approx\frac{0.06 f}{256 \pi^2}\frac{m_i^5}{\bar{M}_{Pl}^4}
\end{align}
where $f\sim (m_i/m_1)^2$ accounts for the enhancement due to the number of different lighter KK-modes that can be produced in this decay. As discussed before, the bulk of the dark matter particles has $m\sim T_r$ in this scenario, and as these are also the ones with the shortest lifetime, we can estimate the lifetime of this population by substituting $T_r$ for $m_i$. These decays lead to the release of an $\mathcal{O}(1)$ fraction of the rest mass and contribute to high-energy cosmic rays in the Universe today. This can be tested by indirect detection experiments. In principle, these gravity-mediated decays will produce all SM states. However, due to the large multiplicity of colour-charged particles, we expect that the decay is approximated well by a generic hadronic decay channel. A combination of different observations compiled by \cite{Ishiwata:2019aet} puts a lower limit on the lifetime of such decays of ultra-heavy dark matter in the range $10^{28}$ to $10^{30}$~seconds depending on the mass. We interpret this as an upper limit on the reheating temperature for a given $m_1$.

\begin{figure}
    \centering
    \includegraphics[width=0.48\linewidth]{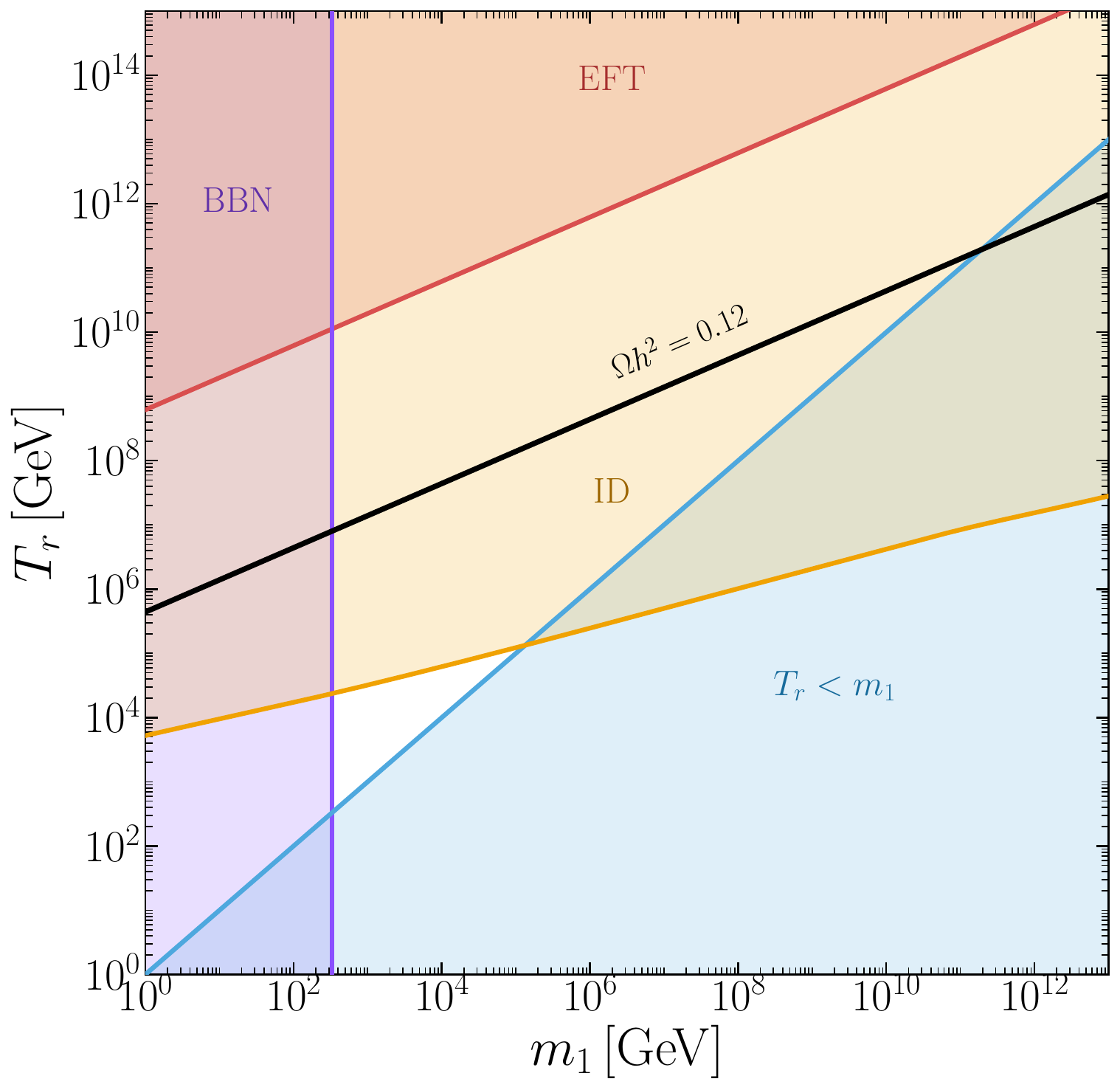}
    \includegraphics[width=0.48\linewidth]{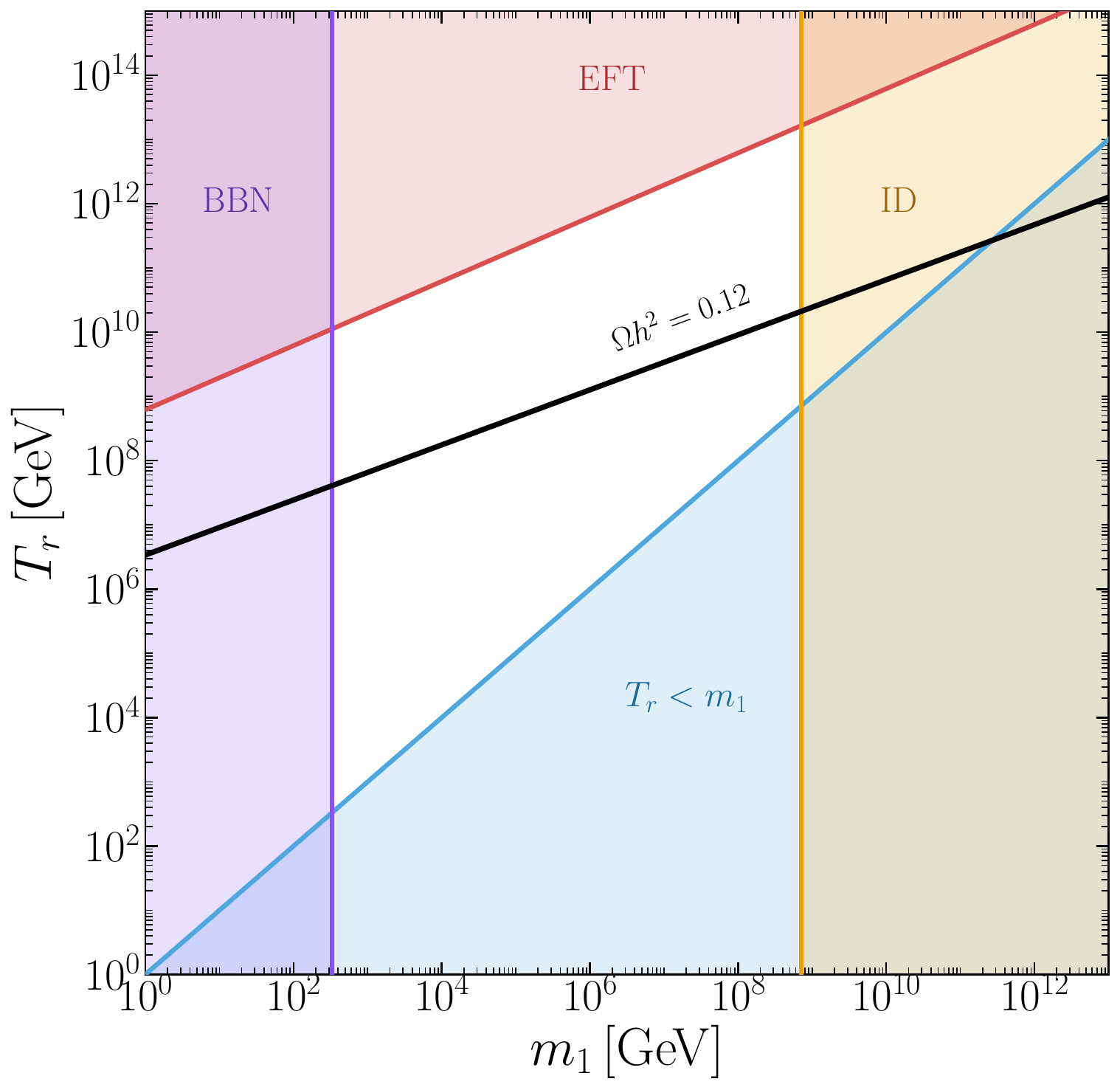}
    \caption{Parameter space of the model as a function of the lightest KK mass $m_1$ and of the reheating temperature $T_r$ for the case in which {\bf (left)} all modes are assumed stable against inter-KK decays and {\bf (right)} KK-number is broken, and all vector modes are assumed to decay to the lightest vector, which is stabilized by the symmetries of the 4D theory. The solid black line indicates the value of the reheating temperature $T_r$ required to produce the correct DM relic density. The red region violates the consistency condition $T_r< M_D$ (EFT). In the blue region $T_r<m_1$ and the relic density is Boltzmann suppressed. The purple region leads to late decays of the KK-gravitons that spoil BBN. Finally, the amber region is excluded by indirect detection~(ID) constraints.
    }
    \label{fig:parameter_space}
\end{figure}

\textit{2) KK-number is broken:} In this case, the more massive vectors decay via a cascade to the lightest KK vector, which is stabilized against further inter-KK tower decays by the structure of the $4D$ theory. Nevertheless, even the lightest vector is not absolutely stable since it can decay into a four SM particle final state via two off-shell KK-gravitons. The kinematics and the matrix element for this process are complex, but due to the high dependence on $m_1$, a precise answer is not required; we believe that a simple estimate based on dimensional analysis should suffice to identify the viable parts of the parameter space. By combining the square matrix element $|\mathcal{M}|^2 \propto m_1^4/M^6_{Pl}$  with the massless four-body phase space \cite{byckling1973particle} we arrive at 
\begin{align}
  \Gamma_{V_{1,4-\text{body}}}\approx \frac{h}{49152 \pi^5}\frac{m_1^7}{\bar{M}_{Pl}^6}
\end{align}
where $h$ is a number that is expected to be of the order of the SM degrees of freedom\footnote{
In contrast to the previous case, there is no parametric enhancement due to the large number of KK-modes since the KK-mass in the denominator of the propagator suppresses their contribution. Summing over all modes thus only leads to an $\mathcal{O}(1)$ correction.}.
Reassuringly, this is $\mathcal{O}(1/\bar{M}_{Pl}^6)$ and thus highly suppressed compared to the width of a KK-graviton of the same mass. However, in order to constitute a viable dark matter candidate, the vectors have to be stable on time scales much longer than the age of the Universe. Imposing the indirect detection limit on the lifetime from  \cite{Ishiwata:2019aet} leads to $m_1\leq 7 \times 10^8$ GeV.

\subsubsection{Results and Discussion}

Combining all the above-mentioned consistency conditions and the observational constraints with our estimations for the DM abundances, we can test which regions of the parameter space of the model allow for a consistent explanation of the dark matter. Our results are summarized in Fig.~\ref{fig:parameter_space}.

The left panel shows the KK conserving case. We choose to span the parameter space by $m_1$, which implicitly fixes $M_D$, and the reheating temperature $T_r$. As expected, the parameter space is limited from above by the requirement $ T_r < M_D$, while the consistency of our relic density computation requires $T_r > m_1$, which sets a lower bound. A lightest Kaluza-Klein mass $m_1\lesssim 300$~GeV is excluded since the decays of the KK-gravitons, which are unavoidably produced as well as the vectors, would spoil BBN. An even stronger upper limit on the mass of the dominant vector modes comes from indirect detection. As the most constraining modes have masses of order of the reheating temperature, the bound is close to horizontal and has very mild dependence on $m_1$. As can be seen, this excludes the part of the parameter space that allows for a production of the full dark matter abundance completely in this case. Therefore, the KK conserving scenario is ruled out in the quoted mass range.

The situation is drastically different for the KK-violating case shown in the right plot of Fig.~\ref{fig:parameter_space}.
While the constraints from BBN, the EFT, and $T_r<m_1$ are the same, the indirect detection ones change. Indeed, in our benchmark scenario, all the heavy modes decay, leaving the indirect detection constraint sensitive only to $m_1$. The constraint then becomes a vertical line bounding from the right. With this modification, the DM compatible region becomes allowed, and solutions do exist within the range $300~\text{GeV}\lesssim m_1 \lesssim 10^9~\text{GeV}$, with a somewhat smaller range of reheating temperature $10^5~\text{GeV}\lesssim T_r \lesssim 10^9~\text{GeV}$. The quoted range follows from the benchmark working assumption of graviphoton number conservation during the decay chain. If more graviphotons were able to survive to the present day, the vertical amber band of the right plot of Fig.~\ref{fig:parameter_space} would start tilting anti-clockwise, reaching the limiting case of the left plot of Fig.~\ref{fig:parameter_space} where all are stable.

\section{Conclusions}
\label{sec:conclusions}

In this work, we have proposed a new dark matter scenario based on the Kaluza-Klein towers of vectors arising in extra-dimensional models with $d\geq2$ compact dimensions. Such a tower of vector modes, the graviphotons, arises unavoidably from the Kaluza-Klein decomposition of the extra-dimensional metric. As these vectors interact with the Standard Model exclusively gravitationally through the other spin-0 and spin-2 components of the extra-dimensional metric, they constitute a natural and minimal dark matter candidate. Interestingly, the full rich particle content of the model is controlled by very few parameters, which can be as small as one for flat universal extra dimensions. For the sake of illustration, we choose such a scenario as a working benchmark. 

The stability of the lightest graviphoton modes is a generic and robust feature of the framework, following from Landau-Yang-type arguments that forbid the decay of a massive vector into two identical massless vectors or into two massless gravitons, and more generally into any pair of identical bosonic states. Together with the absence of a direct coupling to the SM, this leaves only decays suppressed by higher powers of the Planck mass as possible decay channels, making the dark matter candidate naturally long-lived even on cosmological timescales.
Having presented some general considerations for arbitrary $d\geq2$, we specialised to the $d=2$ case as our benchmark scenario. We then derived the full field content and the interactions among the gravitational degrees of freedom and the Standard Model. We argued that the qualitative features of this benchmark, in particular the multi-component dark matter structure and the freeze-in production mechanism, are expected to carry over to larger $d$.

We discussed the cross-sections needed for the cosmological computation of the model, showing how an intricate series of cancellations ensured by the extra-dimensional geometry unitarises them in the high-energy limit, generalising previous results obtained in 5D models.
Building on this, we computed the dark matter yield from freeze-in, separately treating the relativistic, non-relativistic, and intermediate regimes. 
We adopted two benchmark scenarios. The first faithfully follows the model, where KK-conservation induced by translational invariance in the extra dimension makes the entire tower of KK graviphotons extremely long-lived, thus creating a rich multi-component DM scenario. The second one aims to capture a similar model, where KK conservation is broken, making all heavy modes, including graviphotons, unstable. Using an agnostic ansatz for these decays, we find that the graviphotons preferably decay into other graviphotons, thus preserving the total number. This results in a scenario where the number densities of the entire tower of graviphotons get converted entirely into the lightest mode, which then constitutes the total abundance of DM.

 We find that in the KK-conserving case, the entire parameter space where the correct DM abundance is achieved is ruled out by indirect detection constraints. On the other hand, we have also shown that if KK-number conservation is violated, the indirect detection constraints can be lifted, and a large portion of the parameter space becomes available in the range $300~\text{GeV}\lesssim m_1 \lesssim 10^9~\text{GeV}$, with a somewhat smaller range of reheating temperature $10^5~\text{GeV}\lesssim T_r \lesssim 10^9~\text{GeV}$. Such a prediction relies on the assumption that the breaking parameter is small, but efficient enough to make all the heavy vector modes decay. Ultimately, the exact prediction for the available parameter space depends on the type of KK breaking source, and the decay channels that open therein. The analysis of such mechanisms is left for future work.

\section*{Acknowledgments}
A.d.G. thanks M.~Reig and F.~Silvetti for useful discussions.
Part of our numerical analyses were carried out with the help of Claude Code. All physical arguments, calculations, and conclusions were developed and verified by the authors, who take full responsibility for them.
\begin{appendices}
\numberwithin{equation}{section}
\section{High-Energy Limit and Unitarisation}
\label{app:high-energy}

It is well-known that the high-energy behaviour of the amplitudes involving massive graviton final states is plagued by large energy growth in $s$. For the case of the $2\to2$ scattering of gravitons, this can be as large as $\mathcal{O}(s^5)$. This is due to the longitudinal modes of massive gravitons, whose energy growth is given by Eq.~\eqref{pol-tens-2}
\begin{equation}
    \epsilon_0^{\mu\nu} \sim \frac{s}{m^2}\,.
\end{equation} 
The same phenomenon, while less severe, appears in this context when computing final states of massive vectors, as the polarisation tensor of the longitudinal mode of each vector grows as in Eq.~\eqref{pol-tens-1}
\begin{equation}
    \epsilon_0^{\mu} \sim \frac{\sqrt{s}}{m}\,.
\end{equation}
It is of utmost importance for phenomenology to have such behaviour under control.

We now prove that such terms do cancel once all the relevant fields are taken properly into account.
To highlight the geometrical structure of the cancellation and of the extra dimension, we embed the 2D vectors into a 3D vector, such that $\vec{i}=(i_1,i_2,0)^T$ to employ the standard scalar ($\cdot$) and vector product ($\cross$) notation familiar from $3D$. We label the two intermediate states appearing in the $s$-channels as the modes $\sk$ and $\skp$, which we will keep generic. Given the selection rules of Eq.~\eqref{eq:CI-KKselection}, we always have $\psi_\sk a_{\si\sj\sk}=\psi_\skp a_{\si\sj\skp}$, therefore we will omit it in the following. 
We report all amplitudes by helicities. We keep the horizontal lines for the helicity of $V_i$, spanning $+1,0,-1$ from top to bottom; similarly for the columns going from left to right from higher to smaller helicities, if present.

The part of the amplitudes whose energy is faster than $s$ for $HH\to V_i+(\phi_j,h_j)$ are given by
\begin{align}
    &\mathcal{M}_{\phi}^{s>1}= \begin{pmatrix}
        \mathcal{M}_{3/2} s^{3/2}\\
        \mathcal{M}_{3}s^{3}+\mathcal{M}_{2}s^{2}\\
        \mathcal{M}_{3/2} s^{3/2}
    \end{pmatrix}\,, &
    \mathcal{M}_{h}^{s>1}= \begin{pmatrix}
        0&0&\mathcal{M}_{3/2} s^{3/2}&0&0\\
        0&\mathcal{M}_{3/2}s^{3/2} &\mathcal{M}_{3}s^{3}+\mathcal{M}_{2}s^{2} &\mathcal{M}_{3/2}s^{3/2}&0\\
        0&0&\mathcal{M}_{3/2} s^{3/2}&0&0
    \end{pmatrix}\,.
\end{align}
where the amplitudes at each order are not identical for the two final states, but are equally proportional to
\begin{align}
    &\mathcal{M}_{3/2}\propto A_{ijkp}\,,\\ 
    &\nonumber\mathcal{M}_{2}\propto -\frac{3}{2}(i^2+j^2)\left[(k^2+p^2)((\si\cdot\sj)A_{ijkp}-B_{ijkp})-(k^2-p^2)C_{ijkp}\right] \\
    &\qquad\qquad+2(\si\cdot\sk)(\si\cdot\skp)D_{ijkp}+k^2 p^2 \left[B_{ijkp}-3(\si\cdot \sj)A_{ijkp}\right]\cos(2\theta)\\
    &\mathcal{M}_{3}\propto\frac{1}{2} \left[(k^2+p^2)((\si\cdot\sj)A_{ijkp}-B_{ijkp})-(k^2-p^2)C_{ijkp}\right]\,,
\end{align}
and the coefficients are given by:
\begin{align}
    &A_{ijkp}\equiv (\si\cdot\skp) (\si\cross\sk) (i^2 - j^2 + k^2) + 
 (\si\cdot\sk) (\si\cross\skp) (i^2 - j^2 + p^2)\,,\\
    &B_{ijkp}\equiv (\si\cdot\sk) (\si\cdot\skp) (\si\cross\sj) (k^2+p^2-2i^2-2j^2)\,,\\
    &C_{ijkp}\equiv (\si\cdot\sj) (\si\cdot\skp) (\si\cross\sk) (i^2 - j^2 + k^2) - 
  (\si\cdot\sj) (\si\cdot\sk) (\si\cross\skp) (i^2 - j^2 + p^2) 
  \\
  &\nonumber\qquad\qquad-(k^2 - p^2)(\si\cdot\sk) (\si\cdot\skp) (\si\cross\sj) \,,\\
  &D_{ijkp}\equiv 4 (\si\cdot\sj) [p^2 (\si\cdot\sk) (\si\cross\sk)+k^2 (\si\cdot\skp) (\si\cross\skp)]+(k^2+p^2) (\si\cross\sj) (i^4-j^4)\\
   &\nonumber\qquad\qquad-2k^2p^2 (\si\cross\sj)  (i^2-j^2) \,.
\end{align}
By parametrising $\sk=a \si+b\sj$, $\skp=c\si+d\sj$ for some real coefficients $a,b,c,d$, one can show that for the most general $\si,\sj\neq 0$, $A_{ijkp}$ vanishes if and only if
\begin{align}
    &ab+cd=0\,, &&b^2=c^2=d^2=1\,, &&\Longrightarrow&& a,b,c,d=\pm 1\,,
\end{align}
which corresponds only to combinations for which $a_{\si\sj\sk}\neq 0$.
One can check that, given the above conditions, all the other three coefficients also vanish separately, thus cancelling all amplitudes reported above.

For the production of two vectors, $HH\to V_i+V_j$, the situation is similar but slightly more involved. The amplitude can be expanded as
\begin{equation}
    \mathcal{M}_{s>1}^V= \begin{pmatrix}
        0&\mathcal{M}_{3/2} s^{3/2}&0\\
        \mathcal{M}_{3/2}s^{3/2} &\mathcal{M}_{2}s^{2} &\mathcal{M}_{3/2}s^{3/2}\\
        0&\mathcal{M}_{3/2} s^{3/2}&0
    \end{pmatrix}\,.
\end{equation}
The amplitudes $\mathcal{M}_{3/2}$ are all proportional to
\begin{equation}
    \label{eq:caseij-s3/2}\mathcal{M}_{3/2}\propto 2(i^2+j^2)-(k^2+p^2) \propto B_{ijkp}\,,
\end{equation}
which vanishes once the KK conditions are imposed.

The $s^2$ amplitude for vectors is sensibly more complicated. The part of it that depends on the scattering angle $\cos(2\theta)$ is proportional to $(k^2+p^2-2i^2-2j^2)$ as in Eq.~\eqref{eq:caseij-s3/2}, which vanishes once the KK selection rules are imposed. The remaining part of the amplitude is most conveniently organized by first symmetrizing over the relabelings that leave the kinematics invariant. Since the amplitude must be separately symmetric under $i\leftrightarrow j$ and under $k\leftrightarrow p$, we write it as
\begin{equation}
\mathcal{M}_2^V\propto f_{ijkp}+f_{jikp}+f_{ijpk}+f_{jipk}\,,
\label{eq:fsym}
\end{equation}
for a single function $f_{ijkp}$, with the four terms corresponding to the two independent pairwise exchanges. The seed function is built out of two pieces.  Let us define the scalar and vector product by $\si\cdot\sj\equiv S_{ij}$, and $\si\cross\sj\equiv V_{ij}$. The first piece reads
\begin{equation}
a(i,j,k)=i^2\Big[S_{ij}^2\big(S_{ik}^2-S_{jk}^2\big)-S_{ij}V_{ij}\big(S_{ik}V_{ik}+S_{jk}V_{jk}\big)\Big]-(i^2+j^2)\,S_{ik}S_{jk}V_{ik}V_{jk}\,.
\label{eq:aijk}
\end{equation}
 The second piece,
\begin{align}
B(k,p)={}&S_{ij}^2\big(S_{ik}S_{jp}+S_{ip}S_{jk}\big)\big(S_{ik}S_{ip}+S_{jk}S_{jp}\big)\notag\\
&+S_{ij}V_{ij}\Big[-S_{ik}S_{ip}\big(S_{jk}V_{ip}+S_{jp}V_{ik}\big)+S_{ik}S_{jk}S_{jp}V_{jp}+S_{ip}S_{jk}S_{jp}V_{jk}\Big]\notag\\
&-2\,S_{ik}S_{ip}S_{jk}S_{jp}\big(V_{ik}V_{jk}+V_{ip}V_{jp}\big)\,,
\label{eq:Bkk1}
\end{align}
mixes $k$ and $p$ and is already invariant under both $i\leftrightarrow j$ and $k\leftrightarrow p$. The seed function is then
\begin{equation}
f_{ijkp}=p^2\,S_{ip}S_{jp}\,a(i,j,k)-\frac14\,k^2p^2\,B(k,p)\,.
\label{eq:fijkk1}
\end{equation}
The total amplitude vanishes once the KK selection rules are imposed.


\section{Total Abundance Estimation}
\label{app:intermediate-yield}
We derive analytical upper and lower bounds on $S(T_r, N)$ in the regime  $m_1 N \gg T_r \gg m_1$. Starting from the definition
\begin{equation}
    S(T_r, N) = \int_0^{T_r} dT \int_1^N di\, i^2 \int_1^N dj\, j 
    \int_{-\pi/2}^{\pi/2} d\Delta\, \langle\sigma v\rangle_{ij}(\Delta)\,,
\end{equation}
where
\begin{equation}
    \langle\sigma v\rangle_{ij}(\Delta)\, n_{\text{SM,eq}}^2 = 
    \frac{T}{32\pi^4} \int_{(m_i+m_j)^2}^{\infty} ds\;
    \sigma_{\rm eff}(s,m_i,m_j,\Delta)\,\sqrt{s}\,s\,
    K_1\!\left(\frac{\sqrt{s}}{T}\right)\,,
\end{equation}
with $n_{\text{SM,eq}} = T^3/\pi^2$, $m_i = im_1$, $m_j = jm_1$, and
\begin{align}
    \sigma_{\rm eff}(s) =& \frac{\sqrt{(s-(m_i-m_j)^2)(s-(m_i+m_j)^2)}}
    {120\pi\bar{M}_{\rm Pl}^4}\times\\
    &\qquad\qquad\qquad\qquad\nonumber\times
    \left[1 + \frac{14(m_i+m_j)^2m_im_j}{s}
    \left(\frac{1+\cos\Delta}{s_-}+\frac{1-\cos\Delta}{s_+}\right)\right]\,,
\end{align}
where $s_\pm \equiv s - m_i^2 - m_j^2 \mp 2m_im_j\cos\Delta$.
We perform the change of variable $s = (m_i+m_j)^2 + t^2$, $ds = 2t\,dt$,
under which the lower limit maps to $t=0$ and the Källén function factorises as
\begin{equation}
    \sqrt{(s-(m_i-m_j)^2)(s-(m_i+m_j)^2)} = t\sqrt{4m_im_j+t^2}\,,
\end{equation}
while $s_\pm = 2m_im_j(1\mp\cos\Delta)+t^2$, so that
\begin{equation}\label{eq:sigma_t}
    \sigma_{\rm eff} = \frac{t\sqrt{4m_im_j+t^2}}{120\pi\bar{M}_{\rm Pl}^4}
    \left[1 + \frac{14(m_i+m_j)^2m_im_j}{s}\,F(\cos\Delta,t)\right]\,,
\end{equation}
where we have defined
\begin{equation}
    F(\cos\Delta, t) \equiv 
    \frac{1+\cos\Delta}{2m_im_j(1+\cos\Delta)+t^2} + 
    \frac{1-\cos\Delta}{2m_im_j(1-\cos\Delta)+t^2}\,.
\end{equation}
We now bound $F$ from above and below uniformly in $\Delta$. Setting $\partial F/\partial(\cos\Delta)=0$ yields
\begin{equation}
    \frac{t^2}{\left(2m_im_j(1+\cos\Delta)+t^2\right)^2} = 
    \frac{t^2}{\left(2m_im_j(1-\cos\Delta)+t^2\right)^2}\,,
\end{equation}
which is satisfied only at $\cos\Delta = 0$, so $F$ has a unique interior 
extremum. Evaluating at the interior point and at the boundary:
\begin{equation}
    F\big|_{\cos\Delta=0} = \frac{2}{2m_im_j+t^2}\,, \qquad
    F\big|_{\cos\Delta=\pm 1} = \frac{2}{4m_im_j+t^2}\,,
\end{equation}
Since $2m_im_j+t^2 < 4m_im_j+t^2$, the interior point is a maximum 
and the boundary is a minimum, giving the uniform bounds
\begin{equation}
    \frac{2}{4m_im_j+t^2} \leq F(\cos\Delta,t) \leq \frac{2}{2m_im_j+t^2}\,.
\end{equation}
Since both bounds are independent of $\Delta$, the integral over 
$\Delta\in[-\pi/2,\pi/2]$ contributes a factor of $\pi$. We will use these two extreme cases later on.
The thermally averaged cross-section is given by
\begin{equation}
    \langle\sigma v\rangle_{ij} = \frac{1}{32T^5}
    \int_0^\infty 2t\,dt\;\sigma_{\rm eff}\,s^{3/2}\,K_1\!\left(\frac{\sqrt{s}}{T}\right).
\end{equation}
Substituting $u = im_1/T$, $v = jm_1/T$, $\tau = t/T$, the integration measure 
transforms as
\begin{equation}
    i^2\,di\,j\,dj = \frac{T^5}{m_1^5}\,u^2\,du\,v\,dv\,,
\end{equation}
and defining 
$\xi = u+v$ and $r = \sqrt{\xi^2+\tau^2}$, then $s=(rT)^2$ and the thermal rate integral becomes
\begin{equation}
    \int_{-\pi/2}^{\pi/2}d\Delta\,\langle\sigma v\rangle_{ij} = 
    \frac{T^2}{32\cdot 120\,\bar{M}_{\rm Pl}^4}\,\mathcal{I}^\pm(u,v)\,,
\end{equation}
where all factors of $T$ have scaled out and
\begin{equation}
    \mathcal{I}^\pm(u,v) \equiv \int_0^\infty d\tau\;2\tau^2\sqrt{4uv+\tau^2}
    \left[1 + \frac{28\xi^2 uv}{r^2(c^\pm uv+\tau^2)}\right]r^3\,K_1(r)\,,
\end{equation}
with $c^+=2$ for the upper bound and $c^-=4$ for the lower bound.
The mode integrals then give
\begin{equation}
    f(T) \equiv \int_1^N di\,i^2\int_1^N dj\,j\int_{-\pi/2}^{\pi/2}d\Delta\,
    \langle\sigma v\rangle_{ij} = 
    \frac{T^7}{m_1^5\cdot 32\cdot 120\,\bar{M}_{\rm Pl}^4}\,\mathcal{D}^\pm\,,
\end{equation}
where
\begin{equation}
    \mathcal{D}^\pm \equiv \int_0^\infty du\,u^2\int_0^\infty dv\,v\;
    \mathcal{I}^\pm(u,v)
\end{equation}
are pure numerical constants, and the upper limits have been extended to infinity 
since $K_1(r)\sim e^{-r}$ exponentially suppresses modes with $(i+j)m_1\gg T$, rendering $S$ independent of $N$ in the regime of interest. 
The intermediate result is then
\begin{equation}
    \label{eq:f-range}\frac{232\,T^7}{m_1^5\,\bar{M}_{\rm Pl}^4} \leq f(T) \leq \frac{336\,T^7}{m_1^5\,\bar{M}_{\rm Pl}^4}\,.
\end{equation}

Integrating the range of $f(T)$ of Eq.~\eqref{eq:f-range} over $T$ then gives
\begin{equation}
    S(T_r,N) = \int_0^{T_r}dT\,f(T) = 
    \frac{\mathcal{D}^\pm\,T_r^8}{8\cdot 32\cdot 120\,m_1^5\,\bar{M}_{\rm Pl}^4}\,,
\end{equation}
and evaluating $\mathcal{D}^\pm$ numerically yields
\begin{equation}
    \frac{29\,T_r^8}{m_1^5\,\bar{M}_{\rm Pl}^4} \;\leq\; 
    S(T_r, N) \;\leq\; 
    \frac{42\,T_r^8}{m_1^5\,\bar{M}_{\rm Pl}^4}\,.
\end{equation}
Given that most modes lie near $\cos\Delta=0$, we expect the result to be closer to the upper bound.
\section{Small Curvature Randall-Sundrum Model}
\label{app:small-curvature}

In this Section, we study the small curvature limit of the one-dimensional RS~\cite{Randall:1999ee,Randall:1999vf} model. The EOM of the RS model reads
\begin{equation}
    -\partial_y\left[A(y)^4\psi_n'(y)\right]=m_n^2 A(y)^2 \psi_n(y)\,,
\end{equation}
where $A(y)=e^{-ky}$ and $m_n$ is the eigenvalue of the SL problem. In the small curvature limit, $kR\ll 1$, and thus the solution can be expanded as
\begin{align}
    &\psi_n = \psi_{0,n}+(kR)\,\psi_{1,n}+\mathcal{O}[(kR)^2]\,, & &m_n = m_{0,n}+(kR)\,m_{1,n}+\mathcal{O}[(kR)^2]\,.
\end{align}
By solving order by order in $kR$, the linearised even solution reads
\begin{align}
    &\psi_n=\sqrt{2} \cos (n \varphi)+(kR)\frac{2 \left(n^2 (\pi -\varphi) \varphi-3\right) \sin (n \varphi)+6 n \varphi \cos (n \varphi)-\pi  n \cos (n \varphi)}{2 \sqrt{2} n}+\mathcal{O}[(kR)^2]\,,\\
    &m_n=\frac{n}{R}\left(1-kR\frac{\pi}{2}\right)+\mathcal{O}[(kR)^2]\,,
\end{align}
where $\varphi=y/R$.
One can check that the solutions are orthogonal and normalised with respect to the scalar product
\begin{equation}
    \bra{\psi_i}\ket{\psi_j}=\frac{1}{\pi R}\int\limits_{0}^{\pi}dy\, A(y)^2\, \psi_i\psi_j=\delta_{ij}\,.
\end{equation}

Let us now focus on the consequences of such results. Kinematically, nothing has changed since the masses get only multiplied by a universal and $n$-independent factor. In the limit $k\gg 1$, we know that $m_n \approx \pi(n+1/4)ke^{-\pi kR}$, which instead worsens the problem. Therefore, curvature cannot solve the kinematic issues encountered in the main text. However, new non-zero 3-point overlaps appear.
Let us assume $n_{1,2,3}\neq 0$, then we find $a_{n_100}=0$, $a_{n_1 n_20}\propto \delta_{n_1 n_2}$ and
\begin{align}
    & a_{n_1 n_2 n_3}= \frac{k}{\pi}\frac{48 \sqrt{2}\, \left(n_1^2 n_2^2 n_3^2\right) \left[(-1)^{n_1+n_2+n_3}-1\right]}{(n_1+n_2-n_3)^2 (n_1-n_2+n_3)^2 (-n_1+n_2+n_3)^2 (n_1+n_2+n_3)^2}+\mathcal{O}(k^2)\,.
\end{align}
Notice that the first two equations are valid at all orders in $k$, while the third one is valid only at linear order. Interestingly, $a_{n_1n_2n_3}\approx 0$ whenever the sum of the KK-modes is even.
\end{appendices}

\bibliographystyle{BiblioStyle.bst}
\bibliography{Bibliography.bib}

\end{document}